\pdfoutput=1

\documentclass[10pt]{IEEEtran}
\usepackage{amsmath}
\usepackage[T1]{fontenc}% optional T1 font encoding
\usepackage{algorithm}
\usepackage{alphalph}
\usepackage{algorithmic}
\usepackage{listings}
\usepackage{physics,amsmath}
\usepackage{nomencl}
\makenomenclature

\usepackage{amsmath}
\usepackage{graphicx}
\usepackage{color}
\usepackage[cmintegrals]{newtxmath}
\usepackage[caption=false,font=footnotesize]{subfig}
\ifCLASSOPTIONcompsoc
\usepackage[caption=false,font=normalsize,labelfont=sf,textfont=sf]{subfig}
\else
\usepackage[caption=false,font=footnotesize]{subfig}
\fi
\usepackage[colorlinks,citecolor=red,urlcolor=blue,bookmarks=false,hypertexnames=true]{hyperref} 
\usepackage{caption}
\usepackage{lipsum}
\usepackage{bbm}
\usepackage{nameref}
\usepackage{varioref}
\usepackage{hyperref}
\usepackage{setspace}
\newtheorem{theorem}{Theorem}
\newtheorem{lemma}{Lemma}
\newtheorem{proposition}{Proposition}

\begin{document}
%\hsize=6.5in
%\onecolumn
%\linespread{2}
%\doublespacing
\title{\textit{Orbital Angular Momentum}: From State Disentanglement Towards the Secrecy Outage Probability}%: Privacy-Funnel \& Information-Bottleneck}
%\title{Global Unstabilisability and Information Bottlenecks}
\author{Makan~Zamanipour,~\IEEEmembership{Member,~IEEE}
%\author{Author~1~\IEEEmembership{}
\thanks{Manuscript received XX, 20YY; revised X XX, XXXX. Copyright (c) 2015 IEEE. Personal use of this material is permitted. However, permission to use this material for any other purposes must be obtained from the IEEE by sending a request to pubs-permissions@ieee.org. Makan Zamanipour is with Lahijan University, Shaghayegh Street, Po. Box 1616, Lahijan, 44131, Iran, makan.zamanipour.2015@ieee.org. }
%\thanks{}         
}

\maketitle
\markboth{IEEE, VOL. XX, NO. XX, X 2023}%
{Shell \MakeLowercase{\textit{et al.}}: Bare Demo of IEEEtran.cls for Computer Society Journals}
%\IEEEtitleabstractindextext{
\begin{abstract}
\textit{How are we able to fix it so as to theoretically go over it, if the optimality and efficiency of the orbital angular momentum (OAM) multiplexing technique in a multi-input multi-output (MIMO) system fails in practice $-$ something that technically causes a considerable reduction in the maximum achievable value of the transceiver's degree-of-freedom? Is the aforementioned challenging issue theoretically interpretable?} This paper considers the above mentioned challenge in the reconfigurable intelligent surface based systems $-$ as an exemplified system model $-$ and consequently proposes a \textit{Mendelain randomaisation} (MR) technique through which we can further overcome the aforementioned challenge. Towards such end and in this paper, we take into account a Markov chain $\mathcal{G}-\circ-\mathcal{X}-\circ-\mathcal{Y}-\circ-\mathcal{V}$ while $\mathcal{G}$, $\mathcal{X}$, $\mathcal{Y}$ and $\mathcal{V}$ respectively denote \textit{transmit design parameters}, \textit{quality of goods}, \textit{error floor in data detection} and \textit{non-orthogonality set} in the OAM implementation $-$ arisen from other non-optimalities, disturbances and inefficiencies such as the \textit{atmospheric Kolmogorov turbulence}. Furthermore, we consider a \textit{not-necessarily-Gaussian randomisation} method according to which the resultant probability distribution function (PDF) is $2-$D in relation to a definable outage probability in the design problem in relation to $\mathcal{Y}$ $-$ that is, in terms of $\beta_{\mathcal{X}}\theta$ while the randomisation is applied to $\theta$ as the control input which should be optimally derived as the graph-theoretically associating parameter between $\mathcal{X}$ and $\mathcal{Y}$, and $\beta_{\mathcal{X}}$ is the the associating parameter between $\mathcal{G}$ and the exposure $\mathcal{X}$. This PDF is able to actualise a Leader-Follower game $-$ from a Mean-Field game theoretical point of view while the major player is the \textit{reference frame} in the OAM design $-$ achievable in the context of an optimisation problem where the game's PDF realises the control input $\theta$ stem from the PDF of the users, i.e., the OAM states. This kind of design results in removing the non-orthogonalities among the OAM states and in guaranteeing a remarkably more acceptable multiplexing structure. Finally, we extend our work to an active eavesdropper included scenario while we mathematically calculate the least probability of the convergence for the relative algorithm.
\end{abstract}

\begin{IEEEkeywords}
Algorithm's convergence probability, eavesdropper, extinction probability, Hawkes process, irradiance, Karhunen-Loeve, lattice, Mendelian randomisation, mean-field game, OAM states, reference frame, voronoi region.
\end{IEEEkeywords}
\maketitle

\IEEEdisplaynontitleabstractindextext
\IEEEpeerreviewmaketitle

\section{Introduction} 

\IEEEPARstart{I}n 1992 and by Allen et al., the idea of optical vortex was combined to the issue of the orbital angular momentum (OAM) \cite{oam1}-\cite{oam12}. The OAM technique implemented according to the optical vortex is technically defined in an infinite-dimensional Hilbert vector space since it may have a vast number of eigenstates. This point creates an infinite amount of usage and implementation of OAM in the field of communications are enormous \cite{oam7}-\cite{oam11}. 

The information capacity of a single photon can be considerably enhanced if the OAM dimension of photons can be fully utilised for information modulation or multiplexing. This point theoretically guarantees a significantly more acceptable transmission capacity of single-wavelength and single-mode fibers \cite{oam7}-\cite{oam11}. The most advantagouse feature of the OAM technique is the natural orthogonality among different OAM states. 

Inversely, one of the most detrimental knock-on efects is the crosstalk in the context of the ono-orthogonality in reality among different OAM states widely caused by the atmospheric turbulence and the misalignment between the transmitting and receiving antennas $-$ relating to the \textit{reference frame} \cite{oam7}-\cite{oam11}.

Mendelian randomisation (MR) \cite{mend1}-\cite{mend14} is a statistical method to exploite genetic variants as instrumental variables which is able to technically estimate the causal effect of modifiable risk factors on an outcome of interest. A fundamental aim of observational epidemiological interpretation of the MR technique is to identify environmentally modifiable risk factors for disease. This is because of the fact that MR can mimic the effect of some exposure of interest and it is thus not generally open to the usual confounding and reverse causation interpretations. In other words, the primal concept of MR is originates from how to draw inference over environmentally modifiable determinants of disease $-$ which consists of provision of information about alternative biological pathways to the given disease. 

The following criteria are listed for MR compared to other traditional techniques \cite{mend1}-\cite{mend14}:

\begin{itemize}
\item Compared to other traditional and conventional observational studies, the probability of MR being affected by confounding or reverse causation is less.
\item MR is assumption-based, something that shows the necessity of the interpretation over the plausibility of the relative assumptions.
\item For the evaluation of the relevance of the results, other sources of evidence are highly required from a statistical point of view.
\end{itemize}

Among the serious problems in the MR technique \cite{mend1}-\cite{mend14}, it is highly important to note that it may suffer from relatively large direct effects on the outcome whose distribution might have been heavy-tailed $-$ called the \textit{idiosyncratic pleiotropy phenomenon}. Additionally, where owing to privacy concerns individual-level genetic data are not available a case in point, the MR technique's efficiency may degrade. Finally, MR based methods may have the potential of suffering from possible power loss or even biased inference when the chosen genetic variants are in \textit{linkage disequilibrium}. 

\subsection{Motivations and contributions} In this paper, we are interested in responding to the following question: \textit{How can we guarantee highly adequate overcome to parameter uncertainties from a novel point of view? How can we deal with the detrimental knock-on effects of casual co-founders on the overall interpretation problem? How can we cope with the non-orthoganility among the OAM states in practice? What if an active eavesdropper exists? How can we guarantee an extinction probability over the temporal growth density of the eavesdropper's control input?} With regard to the non-complete version of the literature, the expressed question strongly motivate us to find an interesting solution, according to which our contributions are fundamentally described as follows. 
\begin{itemize}
\item \textcolor{red}{\textbf{(\textit{i})}} We go over a multi-input multi-output (MIMO) based reconfigurabe intelligent surface (RIS) based transceiver which includes an OAM technique. 

\item \textcolor{red}{\textbf{(\textit{ii})}} We propose an MR technique in order to further overcome the non-orthogonality among the OAM states wich is wide-spread in practice. Towards this end, we apply a Markov chain $\mathcal{G}-\circ-\mathcal{X}-\circ-\mathcal{Y}-\circ-\mathcal{V}$ while $\mathcal{G}$, $\mathcal{X}$, $\mathcal{Y}$ and $\mathcal{V}$ respectively denote \textit{transmit design parameters}, \textit{quality of goods}, \textit{error floor in data detection} and \textit{non-orthogonality set} in the OAM implementation $-$ originating from some inefficiencies as well as disturbances. 

\item \textcolor{red}{\textbf{(\textit{iii})}} Additionally, we consider a \textit{not-necessarily-Gaussian randomisation} method via which we may define an outage probability in the design problem in relation to $\mathcal{Y}$. 

\item \textcolor{red}{\textbf{(\textit{iv})}} The consequently resultant probability density function (PDF) is able to realise a Leader-Follower game from a Mean-Field game theoretical point of view while the major player is the \textit{reference frame} in the OAM design. The singular value decomposition (SVD) based not-necessarily-Gaussian randomisation procedure cited above is able to be achieved in the context of an optimisation problem where the game's PDF realises the control input arisen from the PDF of the users while the users are the OAM states. 

\item \textcolor{red}{\textbf{(\textit{v})}} Pen-ultimately, we extend our solution to an active eavesdropper included scenario while we define an extinction probability through the time zone over the control input in relation to her. Towards such end, we also define a secrecy outage probability.

\item \textcolor{red}{\textbf{(\textit{vi})}} Finally, we define an algorithm to solve our defined problems and we mathematically prove what probability is required at least for the algorithm to converge.

%\item \textcolor{red}{\textbf{(\textit{v})}} This kind of design results in removing the non-orthogonalities among the OAM states and in guaranteeing a remarkably more acceptable multiplexing structure. 
\end{itemize}

\subsection{General notation} The notations widely used throughout the paper is given in Table \ref{table1}.

\begin{figure}[t]
\centering
\subfloat{\includegraphics[trim={{80 mm} {14 mm} {21 mm} {7mm}},clip,scale=0.45]{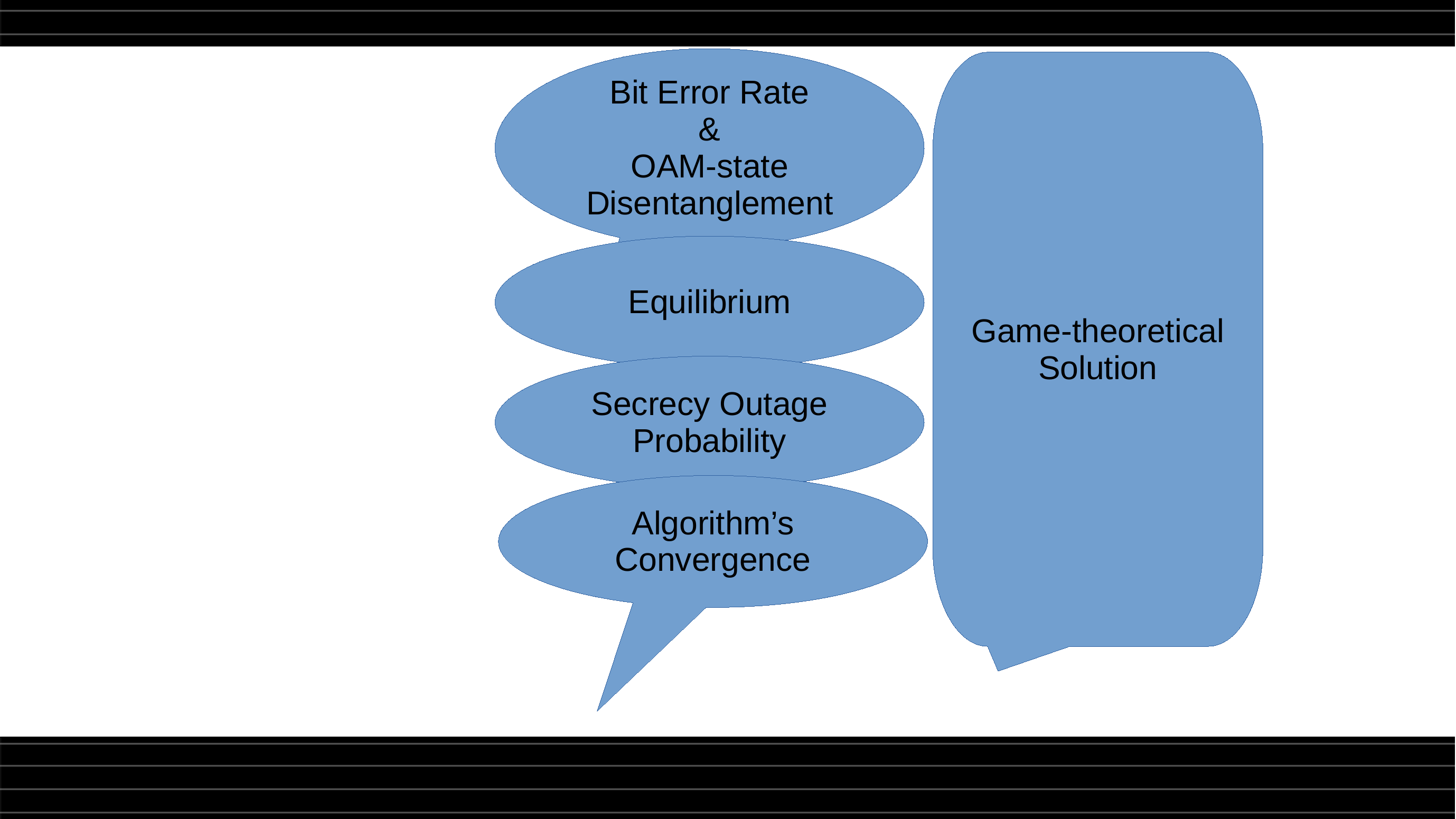}} 
\caption{Flow of the problem-and-solution.} \label{F1}
\end{figure}

\begin{table}[h]
\begin{center}
\captionof{table}{{List of notations.}}
 \label{table1} % for use in \ref{table1} if you want to refer to the table number
\begin{tabular}{p{2.58cm}|p{4.95cm} }%|p{3.46cm}|p{3.84cm}
 \hline      
 \hline
Notation        &  Definition     \\
\hline
    $\mathbb{M}in$   &                   Minimisation              \\
       $\mathcal{I} (\cdot)$      &  Mutual-information      \\
$\stackrel{def}{=} $         &     Is defined as    \\      $\gtrapprox$      &   Greater or approximated to     \\
$\mathscr{p}(\cdot) $         &     Probability densities    \\
     $\varphi_i, i \in \{1, 2\}$      &   Thresholds     \\
$\ell\in \mathscr{L}$ & OAM mode-set index \\
 $L$   & Minimum value: antenna\\
  $\psi_0$     &Transmition-power threshold\\
   $\psi_1$ & Pair of antenna number\\
$\cdot\perp \cdot$ & Orthogonal to\\
$\cdot|\cdot$& Conditionally\\
$||_{\mathscr{kl}}$&Kullback-Leibler Divergence\\
$\vartheta_{i}, i\in \{ \mathcal{X},\mathcal{Y} \}$ &Independent error \\
 $\alpha$&Indirect effect of $\mathcal{G}$ on $\mathcal{Y}$ \\
  $\xi_{\mathcal{X}} $&$\sim \big(0, \partial^2_{\mathcal{X}}\big)$\\
 $\xi_{\mathcal{Y}} $ &$\sim \big(0, \partial^2_{\mathcal{Y}}\big)$\\
  $\mathscr{W}_i, \{ i \in 1, \cdots, 8 \}$&Random-walk Process \\
  $\mathscr{T} \in (0,\infty)$ &Time of the terminal point \\
  $\mathcal{M}$& Game's PDF \\
$\Omega$&Voronoi Region\\
$\Delta$ &Lattice\\
$\Pi$&Hawkes Process\\
$\rho$&Volume-to-noise-ratio\\
$\mathbb{V}\mathscr{ol}$& Volume\\
$(\cdot)_{\mathscr{Eve}}$&Associated to Eavesdropper\\
\hline
 \hline
\end{tabular}
\end{center}
\end{table}

\subsection{Organisation}
The rest of the paper is organised as follows. The system set-up and our main results are given in Section II. Subsequently, the evaluation of the framework and conclusions are given in Sections III and IV, while the proofs are provided in the appendices. The flow of the problem-and-solution is depicted in Fig. \ref{F1}.

\section{Problem formulation \& results}
In this section, we describe the system model, subsequently, we formulate the basis of our problem. 

\subsection{System model: An OAM based RIS assisted MIMO system}

The signal-to-interference-noise-ratio (SINR) of decoding the mode $\ell$ of the OAM set in RIS based transceivers are given as \cite{oam1}-\cite{oam4}
\begin{equation}%\label{eq:1}
\begin{split}
\psi=\mathscr{f}\big(\ell, \psi_0, \psi_1, \psi_2, \psi_3, \psi_4, \psi_5     \big),
\end{split}
\end{equation}
while $\ell\in \mathscr{L}=\{0, 1, \cdots, L-1 \}$ is the OAM mode-set index and $L$ is the minimum value between the transmit and receive antenna, $\psi_0$ stands for the maximum transmition-power threshold, $\psi_1$ denotes the pair of tranmit-receive antenna number in design, $\psi_2$ denotes the angular position set of the antenna elements, $\psi_3$ stands essentially for the radius of the underlying user/equipment, $\psi_4$ is a set for the rotation angle as well as the derivable optimum value for the phase shift for the RIS platform, and finally $\psi_5$ the resulting free-space line-of-sight channel coefficients. The SINR value can be a quality of goods for us in the design problem. 

\textsc{\textbf{Convention 1.}} \textit{Hereinafter and from an information-theoretic point of view, call the random variable}
\begin{itemize}
\item \textit{$\mathcal{G}$ as the set of design parametrs $\big(\ell, \psi_0, \psi_1, \psi_2, \psi_3, \psi_4, \psi_5     \big)$;}
\item \textit{$\mathcal{X}$ as the SINR as the quality of good;}
\item \textit{$\mathcal{Y}$ as the error-floor in data detection; and}
\item \textit{$\mathcal{V}$ as the non-orthogonality set in the OAM multiplexing design.}
\end{itemize}

\subsection{Stochastic association from a graph-theoretical point of view}

\textbf{\textsc{Definition 1: Stochastic association \cite{associated1, associated2}.}} \textit{A finite sequence of random variables $\big( X_1, \cdots, X_n   \big)$ are stochastically associated when for any couple of non-decreasing coordinate-wise and real-valued functions $\mathscr{f}_1\in \mathbb{R}^n$ and $\mathscr{f_2} \in \mathbb{R}^n$ the following holds: $\mathbb{C}ov \Big(\mathscr{f}_1\big( X_1, \cdots, X_n   \big), \mathscr{f}_2\big( X_1, \cdots, X_n   \big)\Big)\ge0$ where $\mathbb{C}ov(\cdot)$ denotes the covariance.}

\textsc{\textbf{Remark 1.}} \textit{The following three essential conditions are satisifed for the Markov chain $\mathcal{G}- \circ- \mathcal{X}-\circ -\mathcal{Y}-\circ -\mathcal{V}$ from an information-theoretic point of view}
\begin{itemize}
\item \textit{(i) $\mathcal{G} \perp \mathcal{V}$ holds, which means that the co-funder set $\mathcal{V}$ is not associated with $\mathcal{G}$;}
\item \textit{(ii) $ \mathcal{Y} \perp \mathcal{G}|(\mathcal{X}, \mathcal{V})$ holds, which means that there exists no association between $\mathcal{G}$ and $\mathcal{Y}$ other than that mediated via $\mathcal{X}$ $-$ that is exclusion-restriction; and}
\item \textit{(iii) the last condition is that any association between $\mathcal{G}$ and $\mathcal{X}$ is not null.}
\end{itemize}

\textsc{\textbf{Goal 1.}} \textit{We aim at removing the effect of $\mathcal{V}$ while we analise $\mathcal{Y}$ with regard to $\mathcal{G}$, chiefly owing to the nature of random data input as the \textit{pleioteropy} remains as a barrier against the validity and reliability of the causal effect estimation. }

\begin{algorithm*}
\caption{{A greedy algorithm to the Problem $\mathscr{P}_2$.}}\label{fff}
\begin{algorithmic}
\STATE \textbf{\textsc{Initialisation.}} 

\textbf{while} $ \mathbb{TRUE}:  \Bigg \vert\hat{\beta}_{\mathcal{X}}(v^+) + \theta \big(v; \hat{\beta}_{\mathcal{X}}(v) \big)- \mathscr{Pr} \Bigg( \theta_{\mathscr{Eve}} \big(v; \hat{\beta}_{\mathcal{X}} \big)=0\bigg \vert \int_0^{v^{-}}\frac{d\theta_{\mathscr{Eve}} \big(v^{\prime}; \hat{\beta}_{\mathcal{X}} \big)}{dv^{\prime}}dv^{\prime}\neq 0\Bigg)\Bigg \vert \le \mathscr{r}_{conv},$ \textbf{do} 

(\textit{i}) Solve $\mathscr{P}_1: 
\begin{cases}
 \partial_t \mathcal{U}+\mathop{{\rm \mathbb{I}nf}}\limits_{\theta (\cdot)} {\rm \; } \big \lbrace \hat{\beta}_{\mathcal{X}} (t) -\bar{\theta}(t) + \mathscr{W}_3\big \rbrace+\mathscr{W}_2=0,\\
\partial _t \mathcal{M}\big(t; \hat{\beta}_{\mathcal{X}}(t)\big)= \partial _{\hat{\beta}_{\mathcal{X}}} \bigg( \theta(t;\hat{\beta}_{\mathcal{X}}) \mathcal{M}\big(t; \hat{\beta}_{\mathcal{X}}(t)\big) \bigg)\\\;\;\;\;\;\;\;\;\;\;\;\;\;\;\;\;\;\;\;\;\;\;\;+\mathscr{W}_4;
\end{cases}$ 

$\;\;\;\;\;\;\;\;\;\;\;\;\;\;\;\;\;\;\;\;\;\;\;\;$and (\textit{ii}) Update. 

\textbf{endwhile} 

\textbf{\textsc{Output.}} 

\textbf{end}

\end{algorithmic}
\end{algorithm*}

According to \textit{Goal 1}, we can information-theoretically write the following
\begin{equation}
\begin{split}
\mathcal{I}\big(      \mathcal{X}; \mathcal{Y}|\mathcal{V}     \big) \ge \mathcal{I}\big(      \mathcal{G}; \mathcal{Y}|\mathcal{V}     \big),
\end{split}
\end{equation}
or
\begin{equation}
\begin{split}
\mathcal{I}\big(      \mathcal{Y}; \mathcal{X}|\mathcal{V}     \big) \ge \mathcal{I}\big(      \mathcal{Y}; \mathcal{G}|\mathcal{V}     \big),
\end{split}
\end{equation}
while 
\begin{equation}
\begin{split}
\mathcal{I}\big(      \mathcal{Y}; \mathcal{G}|\mathcal{V}     \big)=\mathscr{p} \big(\mathcal{Y}, \mathcal{G}, \mathcal{V} \big) \|_{\mathscr{kl}}  \mathscr{p} \big(\mathcal{Y}|\mathcal{V} \big) \underbrace{\mathscr{p} \big(\mathcal{G}|\mathcal{V} \big)}_{\mathscr{p} \big(\mathcal{G} \big)} \mathscr{p} \big(\mathcal{V} \big),
\end{split}
\end{equation}
for which we undoubtedly have the joint probability mass function (PMF)
\begin{equation}
\begin{split}
\mathscr{p}(\mathscr{y}, \mathscr{x}, \mathscr{v}, \mathscr{g})=\mathscr{p}(\mathscr{y}|\mathscr{v}, \mathscr{x})\mathscr{p}(\mathscr{x}|\mathscr{v}, \mathscr{g})\mathscr{p}(\mathscr{v})\mathscr{p}(\mathscr{g}).
\end{split}
\end{equation}

\begin{figure}[t]
\centering
\subfloat{\includegraphics[trim={{73 mm} {85 mm} {118 mm} {14mm}},clip,scale=0.5]{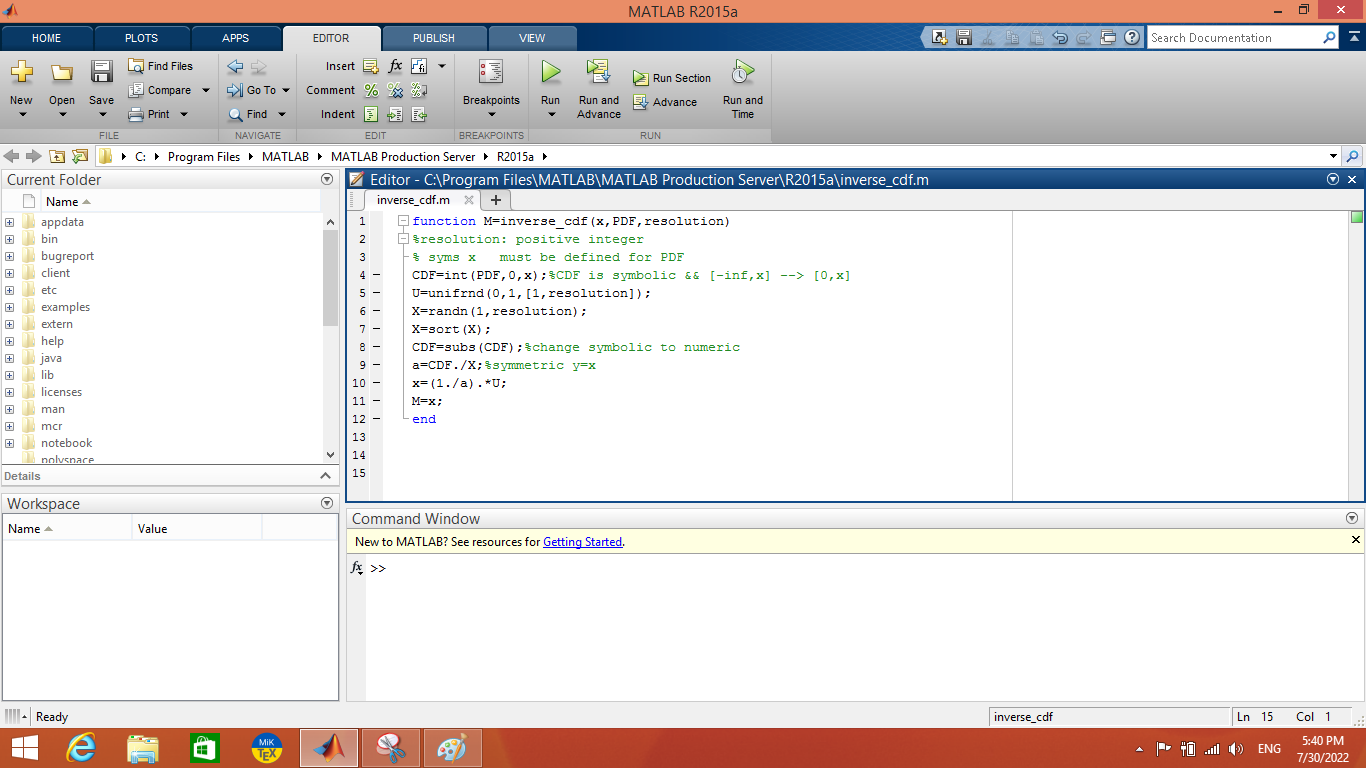}} 
\caption{MATLAB function for the inverse CDF while the PDF is reserved for the game's one.} \label{F2}
\end{figure}

\subsection{MR technique  }

\begin{figure*}[t]
\centering
\subfloat{\includegraphics[trim={{23 mm} {219 mm} {441 mm} {14mm}},clip,scale=0.5]{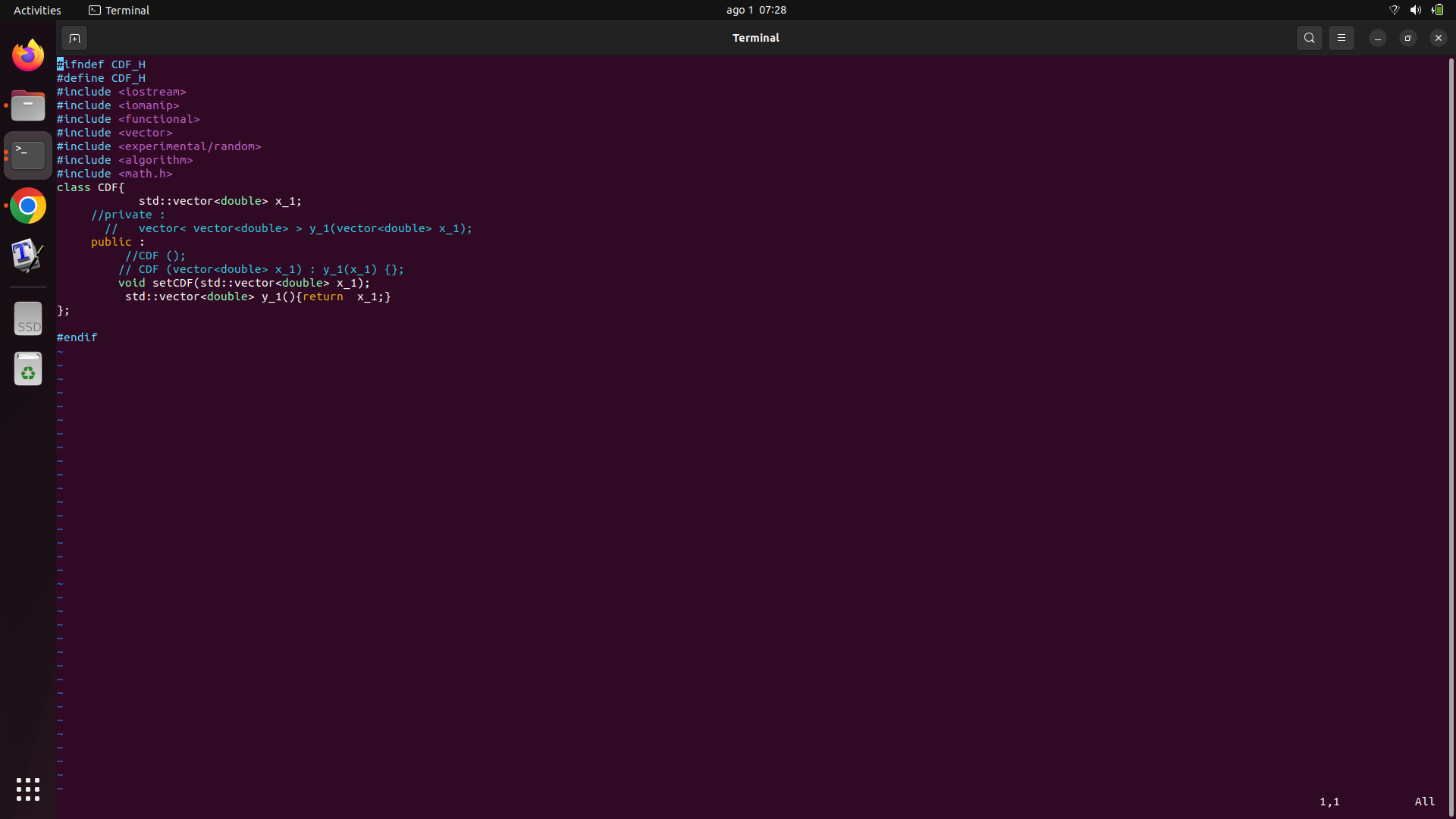}} 
\caption{A $C++$ header $-$ in parallel with the MATLAB syntax.} \label{F3}
\end{figure*}
\begin{figure}[t]
\centering
\subfloat{\includegraphics[trim={{23 mm} {199 mm} {501 mm} {14mm}},clip,scale=0.5]{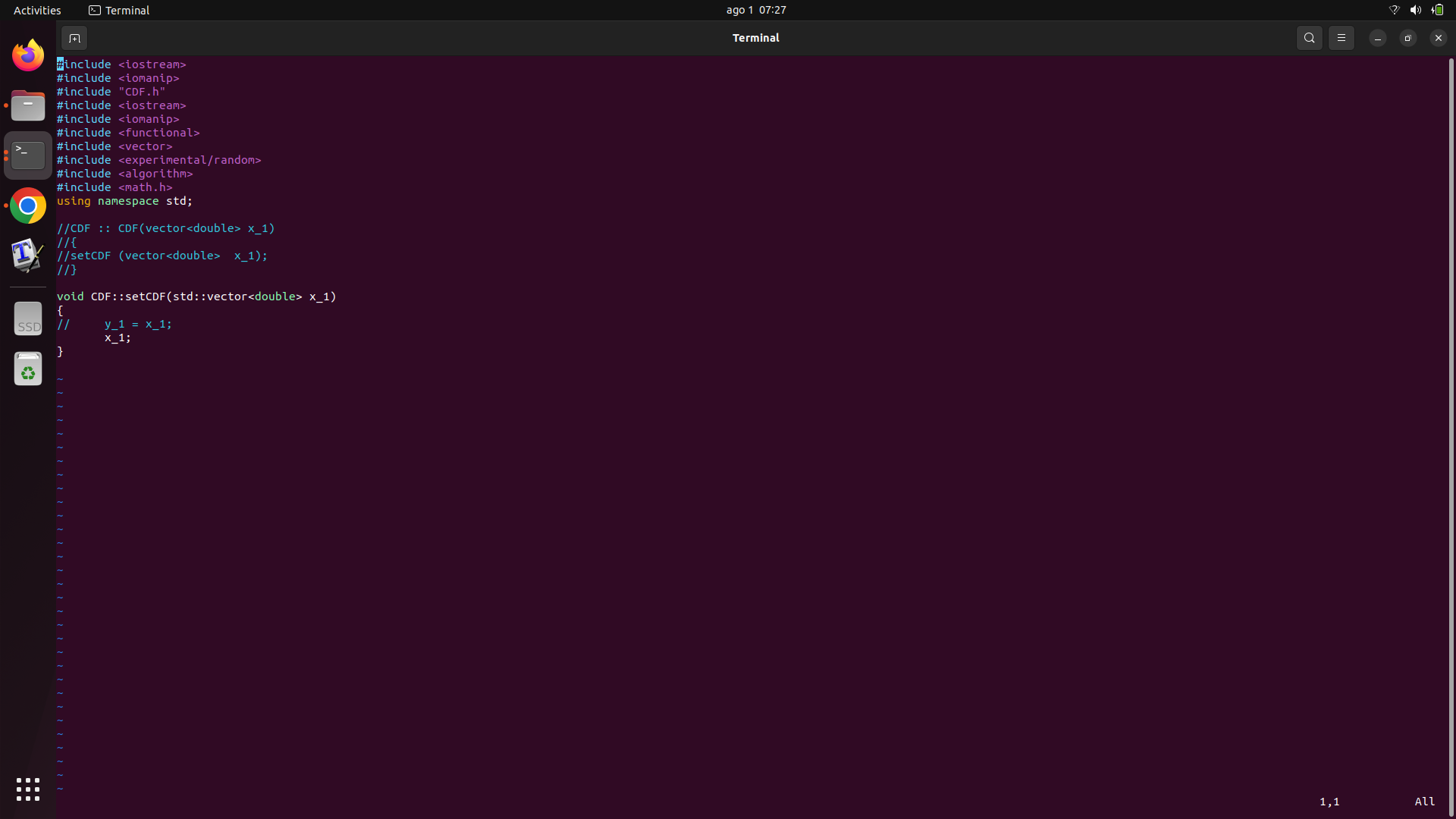}} 
\caption{Auxiliary function in $C++$.} \label{F4}
\end{figure}

The MR technique\footnote{See e.g. \cite{mend1}-\cite{mend14} to see how it works.} is a powerful one in order to infere the precence of the causal effects from the observational data set. One of the most significant beneficial feature of the MR technique in contrast to other pleiotropy robust methods is about the fact that the MR one can be implemented in terms of a multi-variational case. In the following we apply the MR method to solve our main problem. We go

\begin{equation}%\label{eq:1}
\begin{split}
\begin{cases}
\mathcal{X}=\beta^{(0)}_{\mathcal{X}}+\beta_{\mathcal{X}}\mathcal{G}+\gamma_{\mathcal{X}}\mathcal{V}+\vartheta_{\mathcal{X}},\;\;\;\;\\
\mathcal{Y}=\theta_0+\theta\mathcal{X}+\alpha\mathcal{G}+\gamma_{\mathcal{Y}}\mathcal{V}+\vartheta_{\mathcal{Y}},
\end{cases}
\end{split}
\end{equation}
according to which 
\begin{equation}%\label{eq:1}
\begin{split}
\begin{cases}
\hat{\beta}_{\mathcal{X}}=\beta_{\mathcal{X}}+\xi_{\mathcal{X}},\;\;\;\;\;\;\;\;\;\;\;\;\\
\hat{\beta}_{\mathcal{Y}}=\alpha+\beta_{\mathcal{X}}\theta+\xi_{\mathcal{Y}},
\end{cases}
\end{split}
\end{equation}
while$\vartheta_{\mathcal{X}}$ and $\vartheta_{\mathcal{Y}}$ are independent error terms with unit zero means  and unit variances. Also note that $\alpha$ interpretes the indirect effect of $\mathcal{G}$ on $\mathcal{Y}$ and when it is an all-zero matrix then we have no pleiotropy, otherwise, it is a random matrix with given mean and variance. Additionally, $\xi_{\mathcal{X}} \sim \big(0, \partial^2_{\mathcal{X}}\big)$ and $\xi_{\mathcal{Y}} \sim \big(0, \partial^2_{\mathcal{Y}}\big)$ hold.

Now, we aim at technically finding the control input $\theta$ through 
\begin{equation}%\label{eq:1}
\begin{split}
\mathop{{\rm \mathbb{M}in}}\limits_{\theta  } {\rm \; } \frac{1}{\partial^2 _{\mathcal{Y}}}\Big(       \hat{\beta}_{\mathcal{Y}}-\beta_{\mathcal{X}}\theta   \Big)^2,
\end{split}
\end{equation}
or
\begin{equation}%\label{eq:1}
\begin{split}
\mathop{{\rm \mathbb{M}in}}\limits_{\theta_0, \theta  } {\rm \; } \frac{1}{\partial^2 _{\mathcal{Y}}}\Big(       \hat{\beta}_{\mathcal{Y}}-\theta_0-\beta_{\mathcal{X}}\theta   \Big)^2.
\end{split}
\end{equation}

\begin{figure*}[t]
\centering
\subfloat{\includegraphics[trim={{23 mm} {49 mm} {441 mm} {20mm}},clip,scale=0.6]{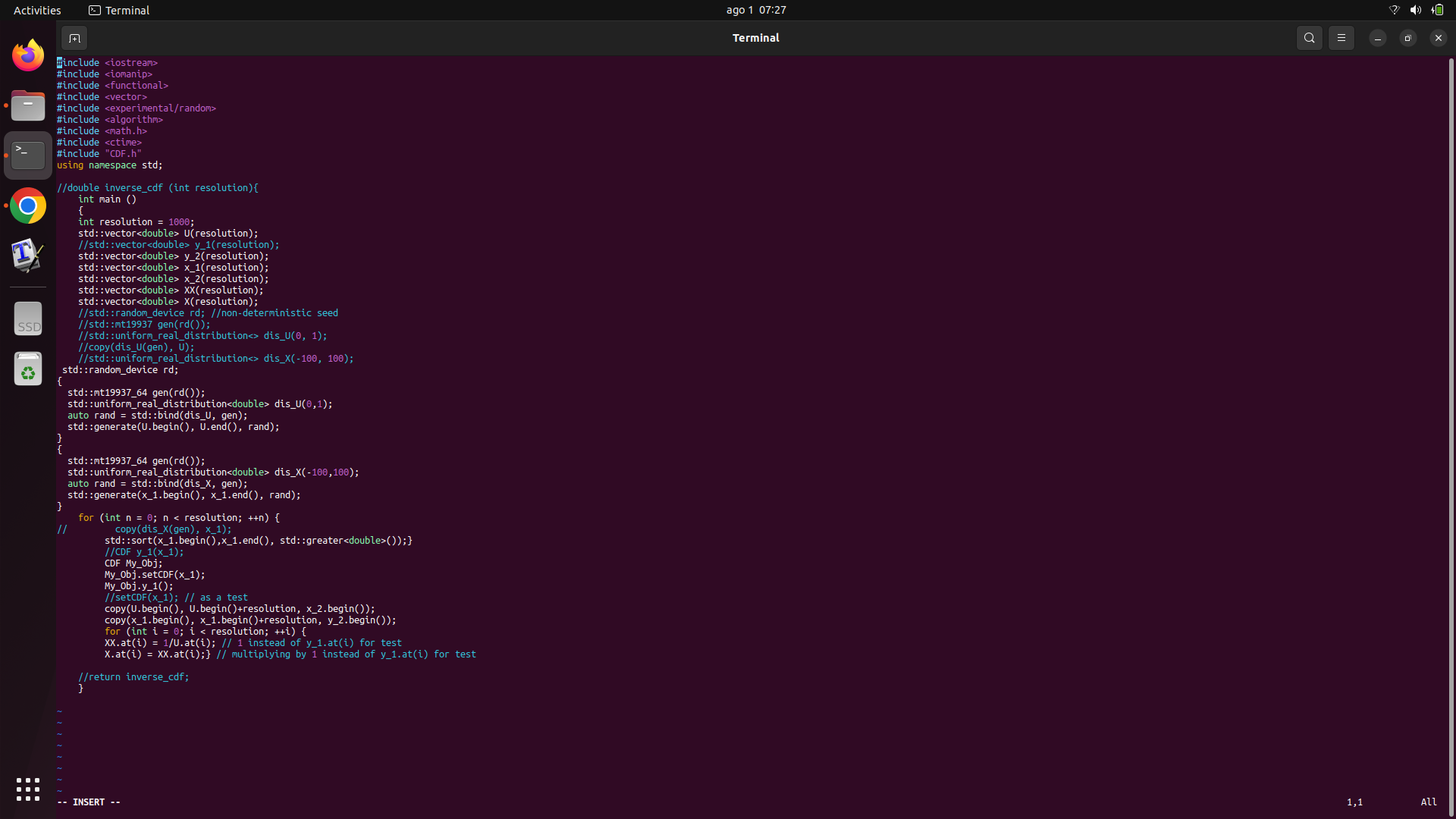}} 
\caption{Main code in $C++$.} \label{F5}
\end{figure*}

\subsection{Mean-Field game-theoretical analysis}

Let $\mathscr{W}_1$ be a random-walk process $-$ noise term and let $\mathscr{T} \in (0,\infty)$ be the time of the terminal point of the control process. Assume the utility function 
\begin{equation}
\begin{split}
 \mathcal{U} \lbrace {t; \hat{\beta}_{\mathcal{X}}(t)} \rbrace= \mathbb{E} \left \{ \int_t^\mathscr{T} \hat{\beta}_{\mathcal{X}}(v) dv \right \},
\end{split}
\end{equation}
over the $2-$D domain $t \in [0, \mathscr{T}] \times \theta(t) \in [\theta(0), \theta(\mathscr{T})]$. Let $\hat{\beta}_{\mathcal{X}}(t=0)=\hat{\beta}_{\mathcal{X}}^{(t=0)}$ and $\hat{\beta}_{\mathcal{X}}(t=\mathscr{T})=\hat{\beta}_{\mathcal{X}}^{(t=\mathscr{T})}$. Considering the control input as $\theta \big(v; \hat{\beta}_{\mathcal{X}}(v) \big)$ and its PDF as $\mathcal{M} \big(v; \hat{\beta}_{\mathcal{X}}(v) \big)\ge 0$ over a complete probability space, the average control input at the time state $v$ is then
\begin{equation}
\begin{split}
 \bar{\theta}(v)=\int_{0}^\infty \theta \big(v; \hat{\beta}_{\mathcal{X}}(v) \big)\mathcal{M} \big(v;\hat{\beta}_{\mathcal{X}}(v) \big)d\hat{\beta}_{\mathcal{X}}.
\end{split}
\end{equation}

Now, the initial equation to reach out the MFG theoretical evaluation as well as the control-law are as follows

\begin{subequations}
\label{eq1}
\begin{align}
\mathop{{\rm min}}\limits_{\theta (\cdot)} {\rm \; }\left \{ \mathcal{U} \left \{ {t_0; \hat{\beta}_{\mathcal{X}}(t_0)} \right \}=\mathbb{E} \left \{ \int_{t_0}^{\mathscr{T}} \hat{\beta}_{\mathcal{X}} dv \right \} \right \}, \label{eq11} \\
d \hat{\beta}_{\mathcal{X}}(v) = - \theta \big(v; \hat{\beta}_{\mathcal{X}}(v) \big)dv + \mathscr{W}_1. \; \; \;\;\;\;\;\;\;\;\;\;\; \label{eq12}  
\end{align}
\end{subequations}

The HJB equation is physically written as 
\begin{equation}
\begin{split}
 \partial_t \mathcal{U}+\mathcal{H}tn+\mathscr{W}_2=0,
\end{split}
\end{equation}
while $\mathscr{W}_2$ is a random-walk process, where the \textit{Hamiltonian} is expressed as 
\begin{equation}
\begin{split}
 \mathcal{H}tn=\mathop{{\rm \mathbb{I}nf}}\limits_{\theta (\cdot)} {\rm \; } \big \lbrace \hat{\beta}_{\mathcal{X}} (t) -\bar{\theta}(t) + \mathscr{W}_3\big \rbrace,
\end{split}
\end{equation}
while $\mathscr{W}_3$ is a random-walk process.

The Bellman principle says that the HJB equation is the necessary and sufficient condition for the value function \cite{c2}, \cite{c3}. Hence, the relative 2-D stochastic partial derivative equation (SPDE) is completed by the FPK equation as 
\begin{equation}
\begin{split}
 \partial _t \mathcal{M}\big(t; \hat{\beta}_{\mathcal{X}}(t)\big)= \partial _{\hat{\beta}_{\mathcal{X}}} \bigg( \theta(t;\hat{\beta}_{\mathcal{X}}) \mathcal{M}\big(t; \hat{\beta}_{\mathcal{X}}(t)\big) \bigg)+\mathscr{W}_4,
\end{split}
\end{equation}
while $\mathscr{W}_4$ is a random-walk process.

\subsection{Outage probability definition}

\textsc{\textbf{Goal 2.}} \textit{We aim at theoretically stochastically associating the PDF of the game $-$ or the control input $\theta$ $-$ , that is, $\mathcal{M}(\cdot)$ and our design parametrs.}

\begin{proposition} \label{P2} \textit{The control input parameter can be related to the OAM states.}\end{proposition} 

\textbf{\textsc{Proof:}} See Appendix \ref{sec:A} for the proof.$\; \; \; \blacksquare$

\begin{proposition} \label{P2} \textit{The upper-bound of the bit-error-rate (BER) is theoretically investigational conditioning on the existence of a Nash-equilibrium.}\end{proposition} 

\textbf{\textsc{Proof:}} See Appendix \ref{sec:B} for the proof.$\; \; \; \blacksquare$

\subsection{How the Scorpion $-$ an active eavesdropper $-$ bites himself}

Immagine an active eavesdropper decides to inteligently intercept the OAM states. The following proposition is valid.

\begin{proposition} \label{P2} \textit{The active eavesdropper's ability to intercept the OAM states is tractable.}\end{proposition} 

\textbf{\textsc{Proof:}} See Appendix \ref{sec:C} for the proof.$\; \; \; \blacksquare$

\begin{theorem} \label{P2} \textit{$\mathscr{P}_2$: The active eavesdropper's ability to intercept the OAM states vanishes over the time zone and the secrecy outage probability is tractable and bounded and the least probability of achieving the gradients has the structure of $1-e^{(\cdot)}$ and calculable. Call this problem as $\mathscr{P}_2$.}\end{theorem} 

\textbf{\textsc{Proof:}} See Appendix \ref{sec:D} for the proof.$\; \; \; \blacksquare$

\subsection{What about the equilibrium?}
In this part, we explore if our game possibly sees any existing Nash-equilibrium or not.

\begin{proposition} \label{P2} \textit{The game is Nash-equilibrium.}\end{proposition} 

\textbf{\textsc{Proof:}} See Appendix \ref{sec:E} for the proof.$\; \; \; \blacksquare$

\begin{proposition} \label{P2} \textit{The probability of Nash-equilibrium creation is tractable and bounded.}\end{proposition} 

\textbf{\textsc{Proof:}} See Appendix \ref{sec:F} for the proof.$\; \; \; \blacksquare$

\subsection{Algorithm to solve Problem $\mathscr{P}_2$}
In this part, we explore if any algorithm to solve $\mathscr{P}_1$ abd $\mathscr{P}_2$ converges and how.

\begin{theorem} \label{P2} \textit{Defining Algorithm \ref{fff}, it converges with probability of at least $\Bigg(1-2 exp \Big(  \frac{-nc_1^{2}}{2}  \Big) \Bigg) \Bigg(  1-2 exp \Big(  \frac{-c_2^{2}}{2 \mathop{{\rm max}}\limits_{n} \vert\vert \mathcal{M}_n \vert\vert_2^2}  \Big) \Bigg)$. Meanwhile, revisiting Algorithm \ref{fff} to solve $\mathscr{P}_2$ according to Theorem 1 and revising the control-law, it converges with probability of at least 
\begin{equation}
\begin{split}
\Bigg(1-2 exp \Big(  \frac{-nc_1^{2}}{2}  \Big) \Bigg) \Bigg(  1-2 exp \Big(  \frac{-c_2^{2}}{2 \mathop{{\rm max}}\limits_{n} \vert\vert \mathcal{M}_n \vert\vert_2^2}  \Big) \Bigg) \cdots \;\;\; \\ \times \Bigg(1-exp \bigg(  \frac{1}{8}\Big( \theta_{\mathscr{Eve}} \big(v^-; \hat{\beta}_{\mathcal{X}} \big)-\theta_{\mathscr{Eve}} \big(v^+; \hat{\beta}_{\mathcal{X}} \big)  \Big)^2  -c_3\bigg)\Bigg).
\end{split}
\end{equation}
}
\end{theorem} 

\textbf{\textsc{Proof:}} See Appendix \ref{sec:G} for the proof.$\; \; \; \blacksquare$

\textbf{\textsc{Remark 2.}} \textit{While $1-\varrho=\Bigg(1-\underbrace{2 exp \Big(  \frac{-nc_1^{2}}{2}  \Big)}_{a} \Bigg) \Bigg(  1-\underbrace{2 exp \Big(  \frac{-c_2^{2}}{2 \mathop{{\rm max}}\limits_{n} \vert\vert \mathcal{M}_n \vert\vert_2^2}  \Big)}_{b} \Bigg)  \Bigg(1-\underbrace{exp \bigg(  \frac{1}{8}\Big( \theta_{\mathscr{Eve}} \big(v^-; \hat{\beta}_{\mathcal{X}} \big)-\theta_{\mathscr{Eve}} \big(v^+; \hat{\beta}_{\mathcal{X}} \big)  \Big)^2  -c_3\bigg)}_{c}\Bigg)$ is the least probability of the convergence, the complexity of the algorithm is of an order of} 
\begin{equation}
\begin{split}
ln(\frac{1}{\varrho}),
\end{split}
\end{equation}
\textit{where $\varrho=1-a-b-c+ab+bc+ac-abc$\footnote{See e.g. \cite{alg1}-\cite{alg5} how.}.}

\section{Numerical results}

Initially opening, we have done our simulations using GNU Octave of version $4.2.2$ on Ubuntu $20.04$ and $22.04$ in parallel with $gcc$ $11.2.0$. 

In Fig. \ref{F2}, a MATLAB function is shown for understanding how to invoke it implementing the randomisation procedure with respect to the PDF of the game described in the paper.

In subsequent of Fig. \ref{F2}, Figs. \ref{F3}, \ref{F4} and Fig. \ref{F5} also show the $c++$ version and the relative headers to invoke it in order to technically apply the procedure cited above $-$  in parallel with the MATLAB syntax.

\begin{figure}[t]
\centering
\subfloat{\includegraphics[trim={{10 mm} {64 mm} {21 mm} {74mm}},clip,scale=0.45]{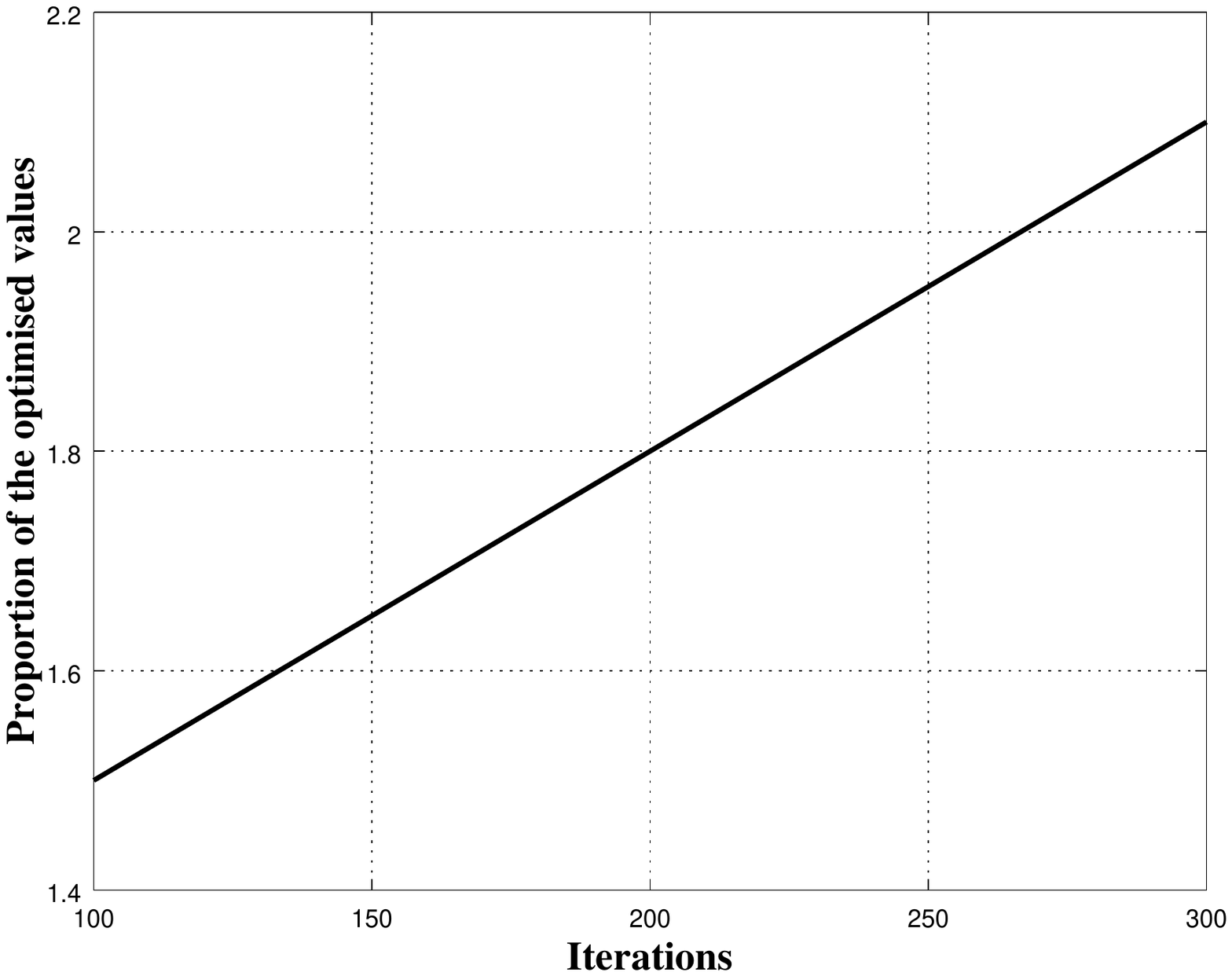}} \\
\subfloat{\includegraphics[trim={{33 mm} {74 mm} {45 mm} {84mm}},clip,scale=0.6]{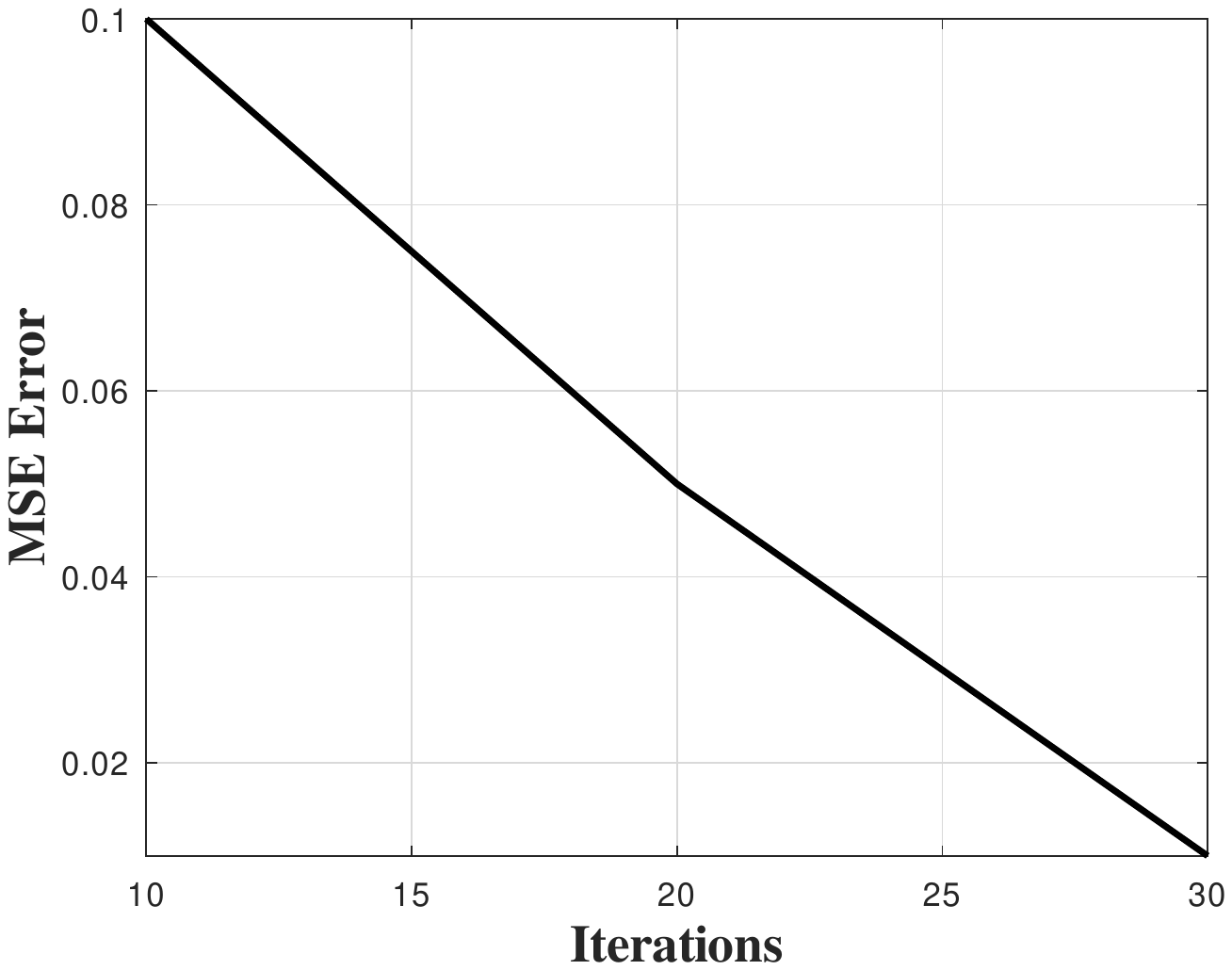}} 
\caption{Optimised values and error versus iterations.} \label{F6}
\end{figure}

In Fig. \ref{F6} and the first subfigure, a proportion of the optimised values obtained from the main optimisation problem is depicted versus the itertaion regime. Consequently, the second subfigure shows a minimum squared error (MSE) is plotted against the iterations.

In Fig. \ref{F7}, we show the trend of the least probability of the convergence for Algorithm \ref{fff} $-$ normalised version $-$ versus the regime of the temporal extinction density in the eavesdropper's control input as $\Big( \theta_{\mathscr{Eve}} \big(v^-; \hat{\beta}_{\mathcal{X}} \big)-\theta_{\mathscr{Eve}} \big(v^+; \hat{\beta}_{\mathcal{X}} \big)  \Big)^2$ $-$ normalised version. 

\begin{figure}[t]
\centering
\subfloat{\includegraphics[trim={{20 mm} {69 mm} {21 mm} {77mm}},clip,scale=0.55]{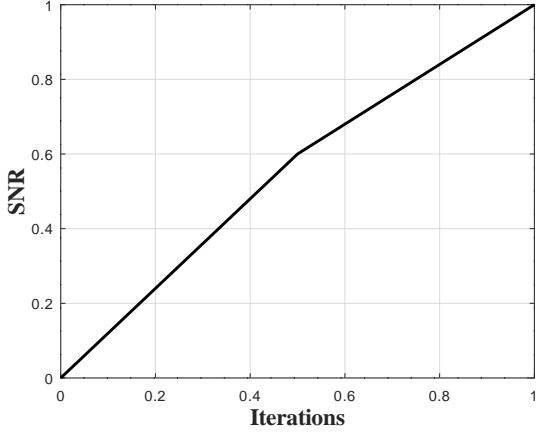}} 
\caption{Trend of the averaged version of SNR $-$ normalised version $-$ versus the regime of the iterations $-$ normalised version.} \label{F8}
\end{figure}

In Fig. \ref{F8}, we show the trend of the averaged version of SNR $-$ normalised version $-$ versus the regime of the iterations $-$ normalised version. 

\begin{figure}[t]
\centering
\subfloat{\includegraphics[trim={{20 mm} {69 mm} {21 mm} {77mm}},clip,scale=0.55]{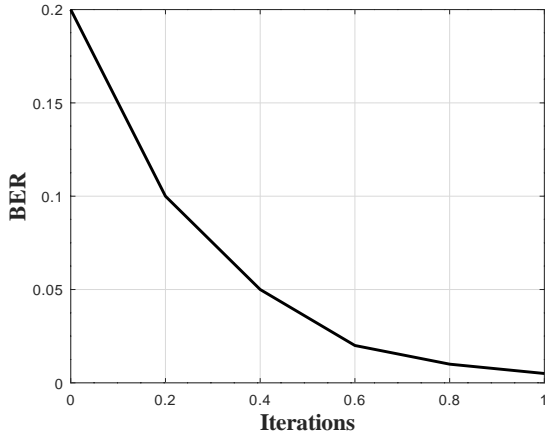}} 
\caption{BER $-$ normalised version $-$ versus the regime of the iterations $-$ normalised version.} \label{F9}
\end{figure}

In Fig. \ref{F9}, we show the trend of BER $-$ normalised version $-$ versus the regime of the iterations $-$ normalised version. 

\section{conclusion}
We technically explored a MIMO based RIS based transceiver which includes an OAM technique. We proposed an MR technique in order to tackle the non-orthogonality among the OAM states. A Markov chain was considered as $\mathcal{G}-\circ-\mathcal{X}-\circ-\mathcal{Y}-\circ-\mathcal{V}$ while $\mathcal{G}$, $\mathcal{X}$, $\mathcal{Y}$ and $\mathcal{V}$ respectively denote \textit{transmit design parameters}, \textit{quality of goods}, \textit{error floor in data detection} and \textit{non-orthogonality set} in the OAM implementation $-$ originating from some inefficiencies as well as disturbances. Furthermore, a \textit{not-necessarily-Gaussian randomisation} method was applied through which we could define an outage probability in the design problem in relation to $\mathcal{Y}$. 

\appendices
\section{Proof of Proposition 1}
\label{sec:A}
The proof is technically initiated through the following outage probability
\begin{equation}
\begin{split}
 \mathbb{P}\mathscr{r} \bigg(   \mathbb{E} \big(    \mathcal{Y}   \big) \le    \varphi_1  \bigg),
\end{split}
\end{equation}
while $\varphi_1$ is a threshold. 

The above outage probability can be re-written as
\begin{equation}
\begin{split}
 \mathbb{P}\mathscr{r} \bigg(   \mathbb{E} \big(    \theta \beta_{\mathcal{X}}\mathcal{G}  \big) \le    \varphi_2  \bigg),
\end{split}
\end{equation}
while $\varphi_2$ is a threshold. Now, since we completely get access to $\mathcal{G}$, we can write\footnote{See e.g. \cite{rand1}-\cite{rand4} to understand the theoretical logic behind of the randomisation principle.}
\begin{equation}
\begin{split}
 \mathcal{G}\stackrel{def}{=}\mathscr{U}_1\Sigma \mathscr{U}_2,
\end{split}
\end{equation}
while we set
\begin{equation}
\begin{split}
\mathscr{PDF}_{\mathscr{U}_1}\stackrel{def}{=}\mathcal{M}(\cdot),
\end{split}
\end{equation}
while for
\begin{equation}
\begin{split}
 \mathbb{P}\mathscr{r} \bigg(   \mathbb{E} \big(    \theta \beta_{\mathcal{X}}\mathcal{G}  \big) \le    \varphi_2  \bigg),
\end{split}
\end{equation}
the underlying PDF is\footnote{See e.g. \cite{cdf1}-\cite{cdf2} to understand the calculation of the PDF of multiplications.}
\begin{equation}
\begin{split}
\frac{1}{|\underbrace{ \theta \beta_{\mathcal{X}}}_{\theta^{\prime}} |} \mathcal{M}(\frac{\cdot}{\theta^{\prime}}),
\end{split}
\end{equation}

The proof is now completed.$\; \; \; \blacksquare$

\section{Proof of Proposition 2}
\label{sec:B}
The proof is as follows, in terms of the following multi-step solution.

\textsc{\textit{Step 1.}}

First of all, take an additional careful look at the following 2-D PDE 
\begin{equation}
\begin{split}
\mathscr{P}_1: 
\begin{cases}
\partial_t \mathcal{U}+\mathop{{\rm \mathbb{I}nf}}\limits_{\theta (\cdot)} {\rm \; } \big \lbrace \hat{\beta}_{\mathcal{X}} (t) -\bar{\theta}(t) + \mathscr{W}_3\big \rbrace+\mathscr{W}_2=0,\\
\partial _t \mathcal{M}\big(t; \hat{\beta}_{\mathcal{X}}(t)\big)= \partial _{\hat{\beta}_{\mathcal{X}}} \bigg( \theta(t;\hat{\beta}_{\mathcal{X}}) \mathcal{M}\big(t; \hat{\beta}_{\mathcal{X}}(t)\big) \bigg)+\mathscr{W}_4.
\end{cases}
\end{split}
\end{equation}

\textsc{\textit{Step 2.}}

Now, each equation of the above-written 2-D PDE has a technical answer $\eta (x,t ) \in \Big \lbrace \mathcal{M}(\cdot), \mathcal{U}(\cdot) \Big \rbrace$ while $x$ traditionally stands for the parameter $ \hat{\beta}_{\mathcal{X}}$, that is, the answer is of an $\eta \big(\hat{\beta}_{\mathcal{X}}, t\big)$.

\textsc{\textit{Step 3.}} 

\textit{Feynman-Kac} formula\footnote{See e.g. \cite{Kac1}-\cite{Kac3} to understand what it says.} technically indicates that the following PDE
\begin{equation}
\begin{split}
\frac{\partial \eta (x, t)}{\partial t}+\mu(x, t) \frac{\partial \eta (x, t)}{\partial x}-\Upsilon(x, t)\eta (x, t)\\+f(x, t)+\mathscr{W}_5=0, t \in [0, \mathcal{T}],\;\;\;\;\;
\end{split}
\end{equation}
with known functions $\Upsilon(\cdot)$ and $f(\cdot)$ has an answer 
\begin{equation}
\begin{split}
 \eta (x, t) =\;\;\;\;\;\;\;\;\;\;\;\;\;\;\;\;\;\;\;\;\;\;\;\;\;\;\;\;\;\;\;\;\;\;\;\;\;\;\;\;\;\;\;\;\;\;\;\;\;\;\;\;\;\;\;\;\;\;\;\;\;\;\;\;\;\;\;\; \\ \mathbb{E} \Bigg(   \int_t^{\mathcal{T}} e^{- \int_t^{t^{\prime}}\Upsilon(x, \tau) d \tau}f(x, t^{\prime})d t^{\prime}+e^{- \int_t^{\mathcal{T}}\Upsilon(x, \tau) d \tau}   \underbrace{\phi(x)}_{\eta (x, \mathcal{T})} \Bigg),
\end{split}
\end{equation}
while $\mathscr{W}_5$ is a random-walk process and for the term $\mu(x, t)$ the following holds
\begin{equation}
\begin{split}
 dx=\mu(x, t)dt+\mathscr{W}_6,
\end{split}
\end{equation}
while $\mathscr{W}_6$ is a random-walk process.

Now, a careful look at the last equation reveals that that is the equivalent of our initial one, that is, $d \hat{\beta}_{\mathcal{X}}(v) = - \theta \big(v; \hat{\beta}_{\mathcal{X}}(v) \big)dv + \mathscr{W}_1$, something that confirms that $\mu(x,t )$ stands for $\theta\big(t; \hat{\beta}_{\mathcal{X}} \big)$.

\textsc{\textit{Step 4.}}

Now, considering $\eta (x,t )$, the \textit{Hermite-Hadamard inequality}\footnote{See e.g. \cite{Hada1}-\cite{Hada3} to understand what it is.} indicates that
\begin{equation}
\begin{split}
\frac{|\partial \Omega|}{|\Omega|} \int_{\Omega \subset \mathbb{R}^{n}}\eta (x,t )dx \le \mathscr{C}_n \int_{\partial \Omega}\eta (y,t )dy,
\end{split}
\end{equation}
while the term $ \mathscr{C}_n$ is a physical constant relying only upon the dimension $n$ if:
\begin{itemize}
\item (\textit{i}) the function $\eta: \Omega \rightarrow \mathbb{R}^{n}$ is positive convex while this positivity-convexity is proven in the next step.
\end{itemize}

 The last equation can now be re-presented as
\begin{equation}
\begin{split}
\Big \vert\frac{\partial \Omega}{\Omega}\Big \vert \int_{\Omega \subset \mathbb{R}^{n}}\eta (x,t )dx \le \mathscr{C}_n \int_{\partial \Omega}\eta (y,t )dy,
\end{split}
\end{equation}
according to the \textit{Cauchy-Schwarz Inequality}.

\textsc{\textit{Step 5: Proof of the convexity of $\mathcal{M}(x, t)$ and $\mathcal{U}(x, t)$.}} We start it by \textit{contradictory}.

\begin{itemize}
\item (\textit{i}) \textit{Vitale's random Brunn-Minkowski inequality}\footnote{Generalised Brunn-Minkowski inequality \cite{1}.} $-$ follows the expression 
\begin{equation}%\label{eq:1}
\begin{split}
\mathbb{V}\mathscr{ol}  \Big \lbrace           \mathbb{E} \big \lbrace  \mathscr{F}\big( f, :\big)  \big \rbrace   \Big \rbrace \ge \mathbb{E}  \Big \lbrace         \mathbb{V}\mathscr{ol}\big \lbrace  \lbrace  \mathscr{F}\big( f, :\big)  \big \rbrace   \Big \rbrace,
\end{split}
\end{equation}
while 
\begin{equation}
\begin{split}
 \mathbb{E} \Bigg( \underbrace{  \int_t^{\mathcal{T}} e^{- \int_t^{t^{\prime}}\Upsilon(x, \tau) d \tau}f(x, t^{\prime})d t^{\prime}+\phi(x)e^{- \int_t^{\mathcal{T}}\Upsilon(x, \tau) d \tau}   }_{\mathscr{F}\big( f, :\big)} \Bigg),
\end{split}
\end{equation}
that is
\begin{equation}
\begin{split}
 \mathbb{E}  \Big \lbrace         \mathbb{V}\mathscr{ol}\big \lbrace  \lbrace  \mathscr{F}\big( f, :\big)  \big \rbrace   \Big \rbrace,
\end{split}
\end{equation}
is upper-bounded by
\begin{equation}
\begin{split}
\mathbb{V}\mathscr{ol}  \Big \lbrace           \mathbb{E} \big \lbrace  \mathscr{F}\big( f, :\big)  \big \rbrace   \Big \rbrace,
\end{split}
\end{equation}

which means that the right-hand side of the equation stand for $\eta (x, t)$ is its upper-bound itself. Thus, it is necessary and sufficient for us to prove the convexity of the aforementioned upper-bound $-$ or find a contradiction $-$ as we know, according to the \textit{Danskin's theorem}, any maximum of a convex function is convex.

\begin{figure}[t]
\centering
\subfloat{\includegraphics[trim={{20 mm} {69 mm} {21 mm} {77mm}},clip,scale=0.55]{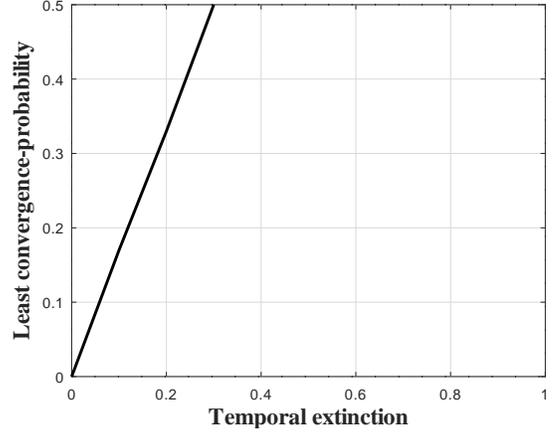}} 
\caption{Trend of the least probability of the convergence for Algorithm \ref{fff} $-$ normalised version $-$ versus the regime of the temporal extinction density in the eavesdropper's control input as $\Big( \theta_{\mathscr{Eve}} \big(v^-; \hat{\beta}_{\mathcal{X}} \big)-\theta_{\mathscr{Eve}} \big(v^+; \hat{\beta}_{\mathcal{X}} \big)  \Big)^2$ $-$ normalised version.} \label{F7}
\end{figure}

%\item (\textit{ii}) We know the term $ e^{- \int_t^{t^{\prime}}\Upsilon(x, \tau) d \tau}$ could be convex even if it is log-concave and no matters if $\Upsilon(x, \tau) $ is concave $-$ as we know if $f$ is concave and $g$ is convex and non-increasing over a univariate domain, then $h ( x ) = g ( f ( x ) ) $ is convex. Nevertheless, it depends on the condition 
%\begin{equation}
%\begin{split}
%\big( \Upsilon(x, \tau) \big)^{\prime \prime}-\Big(\big( \Upsilon(x, \tau) \big)^{\prime }\Big)^2 \ge 0.
%\end{split}
%\end{equation}

\item (\textit{ii}) We know in order to prove that $f(x)$ is convex, we should prove 
\begin{equation}
\begin{split}
f\big(\varpi x_1+(1-\varpi)x_2\big)\le \varpi f(x_1)+(1-\varpi)f(x_2).\\
0 \le \varpi \le 1,\;\;\;\;\;\;\;\;\;\;\;\;\;\;\;\;\;\;\;\;\;\;\;\;\;\;\;\;\;\;\;\;\;\;\;\;\;\;\;\;\;\;\;\;\;\;\;\;\;\;\;\;\;\;
\end{split}
\end{equation}

\item (\textit{iii}) We additionally see that $f(a+b) \ge f(a)+f(b)$ holds due to the expected-value which can be logically interpreted as
\begin{equation}
\begin{split}
f(a)+f(b)=f\big((a+b) . \frac{a}{a+b}\big)+f\big((a+b) .\frac{b}{a+b}\big)\\
\le \frac{a}{a+b}f(a+b)+\frac{b}{a+b}f(a+b),\;\;\;\;\;\\
=f(a+b),\;\;\;\;\;\;\;\;\;\;\;\;\;\;\;\;\;\;\;\;\;\;\;\;\;\;\;\;\;\;\;\;\;\;\;\;\;\;\;
\end{split}
\end{equation}
while we see $-$ if we assign $t_2=0$
\begin{equation}
\begin{split}
f(\varpi t_1)=f\big(\varpi t_1+(1-\varpi).0\big)\;\;\;\;\;\\ \le \varpi f(t_1)+(1-\varpi)f(0) \;\\ \le \varpi f(t_1),\;\;\;\;\;\;\;\;\;\;\;\;\;\;\;\;\;\;\;\;\;\;
\end{split}
\end{equation}
is satisfied, that is, due to $f(0) \le 0$, $f(a+b) \ge f(a)+f(b)$ holds. Now, this is a contradiction regarding the fact that $\mathcal{M}(x, 0)$ and $\mathcal{U}(x, 0)$ are non-negative. The former is from a statistical point of view obvious and the later is due to the fact that the overall system should be supposed to have stability, thus, there may exist some cases in which the aforementioned upper-bound is valid for the scenario of the convexity-positivity cited above $-$ without loss of generality. 

\end{itemize}

\textsc{\textit{Step 6.}}

 The \textit{irradiance} is now tractable with regard to the term $\partial \Omega$. Consequently, the irradiance and its PDF help us in reaching out the signal-to-noise-ratio (SNR) and BER \cite{oam1}-\cite{oam6} $-$ something that is followed in the step below.

\textsc{\textit{Step 7.}} 

Now, the irradiance is as \cite{oam1}-\cite{oam12} the absolute-value of the sume of the OAM beams $-$ \textit{Laguerre Gaussian} ones etc.


Now, let us go in another zone by a transformation application. The key feature of \cite{loeve1}-\cite{loeve3} of the Karhunen-Loeve expansion is that, even though the PDF of the random variables relating to the parameters of this expansion are updated iteratively, the target covariance function is smoothly maintained in terms of: (\textit{i}) orthonormal sets; and (\textit{ii}) an uncorrelated set of random variables.

Let $\big \lbrace \mathcal{X}_t, t \in [a, b] \big \rbrace$ be a q.m. continous second-order process with covariance function $\mathcal{R}(t, s)$. If $\big \lbrace \phi_n  \big \rbrace$ are orthonormal eigenfunctions of the integer operator with kernel $\mathcal{R}(\cdot, \cdot)$ and $\big \lbrace \lambda_n  \big \rbrace$ the corresponding eigenvalues, i.e., 
\begin{equation}
\begin{split}
\int_a^b \mathcal{R}(t, s) \phi_n (s)ds= \lambda_n \phi_n(t), t \in [a, b],
\end{split}
\end{equation}
then we have
\begin{equation}
\begin{split}
 \mathcal{X}(t, w)=\lim_{N \rightarrow \infty} \sum_{n=1}^N  \sqrt{ \lambda_n}a_n(w) \phi_n(t), uniformly \; for \; t \in[a, b], 
\end{split}
\end{equation}
while
\begin{equation}
\begin{split}
 a_n(w)=( \sqrt{ \lambda_n})^{-1}\int_a^b \mathcal{X}(t, w) \phi_n (t)dt, 
\end{split}
\end{equation}
and
\begin{equation}
\begin{split}
 \mathbb{E}[a_ma_n]=\delta_m^n, m \neq n,
\end{split}
\end{equation}
while $\delta$ is \textit{dirac} function.

\textbf{\textsc{Remark 3.}} \textit{Now, we have a linear combination of orthonormally uncorrelated random coefficients for the irradiance the PDF of which is pivotal to calculate BER. Speciffically, $\mathcal{X}(t, w)$ is the transformed version of the irradiance while $a_n$ can be interpreted here as the OAM beams. Thus, the average version of SNR can be reached if we apply Parseval-Theorem for the Karhunen-Loeve transform.}

The proof is now completed.$\; \; \; \blacksquare$

\section{Proof of Proposition 3}
\label{sec:C}
The proof is as follows, in terms of the following \textit{Hawkes process model}.

Let us go from a lattice-theoretic point of view \cite{latt1}-\cite{latt6}. First, re-call the term $\Big \vert\frac{\partial \Omega}{\Omega}\Big \vert$. This term can be considered as the \textit{effective radius} for the Lattice $\Lambda$ in the main volume $\Omega$, call \textit{Voronoi region} \footnote{See e.g. \cite{latt1}-\cite{latt6} to what Voronoi region is.}, in which we are supposed to have the lattice blocks and points. Now, take into account some lattice blocks with random and correlated intervals that have arrivals in order of occurrence. We denote 
\begin{equation}
\begin{split}
\Pi=\big(  \pi^{(t)}_0, \cdots, \pi^{(t)}_n  \big), t \in [0, \mathscr{T}],
\end{split}
\end{equation}
as an $n-$dimensional mutually-exciting point-process modeled by a multi-variate Hawkes process. 

\textbf{\textsc{Remark 4.}} \textit{Since the point processes do not only self-excite, but in general also inteact with each other as well, that is, cross-excite, thus, an $n-$dimensional mutually-exciting point-process is used as a multi-variate Hawkes process\footnote{See e.g. \cite{Haw1}-\cite{Haw2} to understand Hawkes processes.}.} 

The intensity process is now denoted by 
\begin{equation}
\begin{split}
\Delta=\big( \delta^{(t)}_0, \cdots, \delta^{(t)}_n  \big), 
\;\;\;\;\;\;\;\;\;\;\;\;\;\;\;\;\;\;\;\;\;\;\;\;\;\;\;\;\;\;\;\;\;\;\;\;\;\;\;\;\;\;\;\;\;\;\;\; \; \;  \\
  \delta^{(t)}_{j \neq i}=\delta_{j \neq i} \Big(  t \big \vert  \big \lbrace \mathscr{H}_i(t) \big \rbrace  \Big)=  \frac{\mathbb{E}  \big \lbrace  d \Pi^{(t)}   \big \rbrace  }{dt}  =  \;\;\;\;\;\;\;\;\;\;\;\;\;\;\;\;\;\;\;\;\; \; \; \; \\
\gamma_i+      \sum_{i=1}^{m} \int_0^t  \zeta(t-v)\Pi^{(t)} dv, t \in [0, \mathscr{T}], i \in \{0, \cdots, n \},
\end{split}
\end{equation}
while $\gamma_i \ge 0$ is \textit{background intensity}\footnote{See e.g. \cite{Haw1}-\cite{Haw2} to understand what it is.}, $\zeta: \mathbb{R}^{n, +}\rightarrow \mathbb{R}^{n, +}$ is called the \textit{triggering function} that physically actaulises the excitatory impact on the current event arisen from the past one(s), and $\mathscr{H}_i(t)$ denotes the event history associated to the $i-$th process at the time instant $t$. 

The initial control-law $d \hat{\beta}_{\mathcal{X}}(v) = - \theta \big(v; \hat{\beta}_{\mathcal{X}}(v) \big)dv + \mathscr{W}_1$ can now be consequently re-defined as
\begin{equation}
\begin{split}
d \hat{\beta}_{\mathcal{X}}(v) = - \theta \big(v; \hat{\beta}_{\mathcal{X}}(v) \big)dv\;\;\;\;\;\;\;\;\;\;\;\;\;\;\;\;\;\;\;\;\;\;\;\;\;\;\\ +\bigg(\int_0^v \Delta^{(v^{\prime})}dv^{\prime}-\Pi^{(v^{\prime})}\bigg)dv+ \mathscr{W}_7,
\end{split}
\end{equation}
while $ \mathscr{W}_7$ is a random-walk process and the term $\int_0^v \Delta^{(v^{\prime})}dv^{\prime}-\Pi^{(v^{\prime})}$ is called the \textit{compensation process}.

The proof is now completed.$\; \; \; \blacksquare$

\section{Proof of Theorem 1}
\label{sec:D}
The proof is as follows, in terms of the following \textit{Galton-Watson process model}\footnote{See e.g. \cite{Galton} to understand what it is.} and the following multi-step solution.

\textsc{\textit{Step 1.}}

For the model of a Galton-Watson process, we need an \textit{Extinction probability}. Towards such end, let us re-call the Lattice $\Lambda$. A metric named \textit{volume-to-noise-ratio}\footnote{See e.g. \cite{latt1}-\cite{latt6} to understand it.} can be defined for the Lattice $\Lambda$ as 
\begin{equation}
\begin{split}
\rho \big(  \Lambda, \mathscr{P}^{(\mathscr{e})}_{ \pi^{\prime}}  \big)\stackrel{def}{=}\frac{\big\vert \Omega \big\vert^{\frac{cte}{n}}}{\partial^2_{ \pi^{\prime}}},
\end{split}
\end{equation}
while $\partial^2_{ \pi^{\prime}}$ is the variance of the point process $\pi^{\prime}$ whith the error probability
\begin{equation}
\begin{split}
 \mathscr{P}^{(\mathscr{e})}_{ \pi^{\prime}} \ge \mathscr{Pr} \big \lbrace \pi^{\prime}  \notin \Omega \big \rbrace,
\end{split}
\end{equation}

Thus, the initial control-law $d \hat{\beta}_{\mathcal{X}}(v) = - \theta \big(v; \hat{\beta}_{\mathcal{X}}(v) \big)dv + \mathscr{W}_1$ can now be re-defined as
\begin{equation}
\begin{split}
d \hat{\beta}_{\mathcal{X}}(v) = - \theta \big(v; \hat{\beta}_{\mathcal{X}}(v) \big)dv+\;\;\;\;\;\;\;\;\;\;\;\;\;\;\;\;\;\;\;\;\;\;\;\;\;\;\;\;\;\;\;\;\;\;\;\;\;\;\;\;\;\;\;\; \\ \mathscr{Pr} \Bigg( \theta_{\mathscr{Eve}} \big(v; \hat{\beta}_{\mathcal{X}} \big)=0\bigg \vert \int_0^{v^{-}}\frac{d\theta_{\mathscr{Eve}} \big(v^{\prime}; \hat{\beta}_{\mathcal{X}} \big)}{dv^{\prime}}dv^{\prime}\neq 0\Bigg)dv+ \mathscr{W}_8,
\end{split}
\end{equation}
while $ \mathscr{W}_8$ is a random-walk process, $ \mathscr{Pr} \Big( \theta_{\mathscr{Eve}} \big(v; \hat{\beta}_{\mathcal{X}} \big)=0\Big)$ stands for the probability of extinction over the control input in relation to the eavesdropper, and the term $(\cdot)_{\mathscr{Eve}}$ stands for the eavesdropper.

\textsc{\textit{Step 2.}}

The secrecy outage probability can be defined as $\mathscr{Pr} \Big(   \mathscr{Out}   \Big)$ which is a function of the eavesdropper's probability of extinction $ \mathscr{Pr} \Big( \theta_{\mathscr{Eve}} \big(v; \hat{\beta}_{\mathcal{X}} \big)=0\Big)$. We give an example here in order to show if the probability can be tractable or not. Re-call the PDF $\frac{1}{|\theta^{\prime}|} \mathcal{M}(\frac{\cdot}{\theta^{\prime}})$ achieved in Appendix \ref{sec:A} of which the eavesdropper's probability of extinction $ \mathscr{Pr} \Big( \theta_{\mathscr{Eve}} \big(v; \hat{\beta}_{\mathcal{X}} \big)=0\Big)$ can be interpreted. One example of the PDF $\frac{1}{|\theta^{\prime}|} \mathcal{M}(\frac{\cdot}{\theta^{\prime}})$ can be $\frac{c_1c_2}{c_3}log \frac{2}{c_3}$. Thus, one can see that with probability at least $1-exp \Big(  \frac{nc_3}{c_1} \Big)$, we have\footnote{See e.g. \cite{concentraion1}-\cite{concentraion11} for concentration inequalities.}
\begin{equation}
\begin{split}
\sum_{i=1}^n \mathscr{Pr}\big \lbrace \pi_i^{\prime}  \in \Omega \big \rbrace \ge nc_3,
\end{split}
\end{equation}

only regarding the aforementioned PDF while $n$ stands for the dimension and $c_1, c_2, c_3$ are constant values.

\textsc{\textit{Step 3.}}

\begin{lemma} \label{P2} \textit{The term $\mathscr{Pr}\big \lbrace \pi_i^{\prime}  \in \Omega \big \rbrace$ is theoretically bounded from above.}\end{lemma} 

\textit{Proof:} The proof of \textit{Lemma 1} is as follows. We technically know\footnote{See e.g. \cite{concentraion1}-\cite{concentraion11} for concentration inequalities.}
\begin{equation}
\begin{split}
\mathscr{Pr}\big \lbrace \pi_i^{\prime}  \in \Omega \big \rbrace=\sum_{\mathscr{d}_0 \in \pi_i^{\prime}  \cap \Omega} \mathscr{Pr}\big \lbrace \pi_i^{\prime}  = \mathscr{d}_0 \big \rbrace\;\;\;\;\;\;\;\;\;\\  \le 
\sum_{\mathscr{d}_0 \in \pi_i^{\prime}  \cap \Omega} e^{-n \mathscr{D}_{Kl} (\mathscr{d}_0 \vert \cdot )}\;\;\;\;\;\;\;\;\;\;\;
\\
 \le \underbrace{ \big \vert \pi_i^{\prime}  \cap \Omega \big \vert e^{-n \mathop{{\rm \mathbb{I}nf}}\limits_{\mathscr{d}_0 \in \pi_i^{\prime}  \cap \Omega} \mathscr{D}_{Kl} (\mathscr{d}_0 \vert \cdot )} }_{\varsigma_1},
\end{split}
\end{equation}
which can be the upper-bound for the \textit{Sanov theorem} for the large devitaion principle while we call it $\varsigma_1$ $-$ some thing that completes the proof for \textit{Lemma 1}.$\; \; \; \blacksquare$

\textsc{\textit{Step 4.}}

\begin{lemma} \label{P2} \textit{The term $\mathscr{Pr}\big \lbrace \pi_i^{\prime}  \in \Omega \big \rbrace$ is theoretically bounded from below.}\end{lemma} 

\textit{Proof:} The proof of \textit{Lemma 2} is as follows. We technically know\footnote{See e.g. \cite{concentraion1}-\cite{concentraion11} for concentration inequalities.} according to the \textit{exponential tightness theorem} that there exists a $\varsigma_2 \in [0, \infty)$ while for $\Omega_0 \subseteq \Omega$
\begin{equation}
\begin{split}
\lim_{n \rightarrow \infty}\mathop{{\rm \mathbb{S}up}}\limits_{\cdot}\mathscr{Pr}\big \lbrace \pi_i^{\prime}  \in \Omega_0 \big \rbrace \ge \varsigma_2,
\end{split}
\end{equation}
which means that the law $\pi_i^{\prime} $ is exponentially tight and this completes the proof of \textit{Lemma 2}.$\; \; \; \blacksquare$

\textsc{\textit{Step 5.}}

\begin{lemma} \label{P2} \textit{For the control law $d \hat{\beta}_{\mathcal{X}}(v) = - \theta \big(v; \hat{\beta}_{\mathcal{X}}(v) \big)dv+\;\;\;\;\;\;\;\;\;\;\;\;\;\;\;\;\;\;\;\;\;\;\;\;\;\;\;\;\;\;\;\;\;\;\;\;\;\;\;\;\;\;\;\; \\ \mathscr{Pr} \Bigg( \theta_{\mathscr{Eve}} \big(v; \hat{\beta}_{\mathcal{X}} \big)=0\bigg \vert \int_0^{v^{-}}\frac{d\theta_{\mathscr{Eve}} \big(v^{\prime}; \hat{\beta}_{\mathcal{X}} \big)}{dv^{\prime}}dv^{\prime}\neq 0\Bigg)dv+ \mathscr{W}_8$, with probability of at least $1-2e^{\frac{-2c_0^{2}}{\sum_{i=1}^n\Big(\varsigma^{(k)}_2-\varsigma^{(k)}_1\Big)^2}, \; c_0 \ge 0}$ the gradients are not obtained while $c_0$ is a constant.}\end{lemma} 

\textit{Proof:} The proof of \textit{Lemma 3} is as follows. We technically know \textit{Azuma-Hoeffding's inequality} indicates that\footnote{See e.g. \cite{concentraion1}-\cite{concentraion11} for concentration inequalities.} if the condition $a_k \le X_k -X_{k-1} \le b_k$ holds for the super-martingale $\big \lbrace X_n \big \rbrace_{n=0}^{\infty}$ with regard to the constants $a_k$ and $b_k$, $\mathscr{Pr}\Big \lbrace \big \vert X_n-X_0 \big \vert >c_0 \Big \rbrace \le 2e^{\frac{-2c_0^{2}}{\sum_{i=1}^n(b_k-a_k)^2}, \; c_0 \ge 0}$, thus, we have $\mathscr{Pr}\Big \lbrace \big \vert \hat{\beta}_{\mathcal{X}}(v)-\hat{\beta}_{\mathcal{X}}(0) \big \vert <c_0 \Big \rbrace \ge 1-2e^{\frac{-2c_0^{2}}{\sum_{i=1}^n\Big(\varsigma^{(k)}_2-\varsigma^{(k)}_1\Big)^2}, \; c_0 \ge 0}$ while $\varsigma_1$ and $\varsigma_2$ are obtained in respectively \textit{Lemma 1} and \textit{Lemma 2} which stand for the right-hand-side of the control-law expressed above, and now we see this converstaion completes the proof of \textit{Lemma 3}.$\; \; \; \blacksquare$

The proof is now completed.$\; \; \; \blacksquare$

\section{Proof of Proposition 4}
\label{sec:E}
The proof is as follows, in terms of the following muti-step solution.

\textsc{\textit{Step 1.}}

\textsc{\textbf{Definition 2: Simply connected space\footnote{See e.g. \cite{curvature1}-\cite{curvature5}.}.}} \textit{A topological space $Psi$ is called simply connected if it is path-connected and any loop in there defined by $f: \mathcal{S}^1 \rightarrow \Psi$ can be contracted/shrunk to a point.}

\textsc{\textbf{Definition 3: Sectional curvature\footnote{See e.g. \cite{curvature1}-\cite{curvature5}.}.}} \textit{Given two linearly independent tangent vectors at the same point as $\nu_1$ and $\nu_2$, and with regard to }
\begin{equation}
\begin{split}
\mathcal{R}(\nu_1, \nu_2)\nu_3=\nabla_{\nu_1}\nabla_{\nu_2}\nu_3-\nabla_{\nu_2}\nabla_{\nu_1}\nu_3-\nabla_{[\nu_1, \nu_2]}\nu_3,
\end{split}
\end{equation}
{as the Riemann curvature tensor, the sectional curvature can be defined as}
\begin{equation}
\begin{split}
\mathcal{K}(\nu_1, \nu_2)=\frac{<\mathcal{R}(\nu_1, \nu_2)\nu_2, \nu_1>}{<\nu_1, \nu_1><\nu_2, \nu_2>-<\nu_1, \nu_2>^2}.
\end{split}
\end{equation}

\textsc{\textbf{Definition 4: Ricci curvature\footnote{See e.g. \cite{curvature1}-\cite{curvature5}.}.}} \textit{For the matrix $y$ for the main Riemannian manifold, if the Jacobian matrix is $\mathcal{J}$, we have the following for the Ricci curvature}
\begin{equation}
\begin{split}
\bar{\mathscr{R}}=\mathcal{J}^T \big ( \mathscr{R} \circ y\big )\mathcal{J},
\end{split}
\end{equation}
 \textit{while $(\cdot)^T$ stands for the transpose.}

\textsc{\textit{Step 2: Cheng eigen-value comparison theorem\footnote{See e.g. \cite{cheng1}, \cite{cheng2} to understand it.}.}} 

First, let $\mathscr{M}$ be a Riemannian manifold of the dimension $n$, and denote $\mathscr{B}_{\mathscr{M}}(\mathscr{p}, \mathscr{r})$ a geodesic ball centred at $\mathscr{p}$ with radius $\mathscr{r}$ less than the \textit{injectivity radius} of $\mathscr{p} \in \mathscr{M}$. Now, for a real number $k$, denote $\mathscr{N}(k)$ the simply connected space form of dimension $n$ and constant sectional curvature $\mathscr{K}_{\mathscr{M}}$ which is bounded from above to $k$. Now, \textit{Cheng eigen-value comparison theorem} indicates that
\begin{equation}
\begin{split}
 \lambda_1 \big(\mathscr{B}_{\mathscr{N}(k)}(\mathscr{r}) \big)\le \lambda_1 \big(\mathscr{B}_{\mathscr{M}}(\mathscr{p}, \mathscr{r}) \big),
\end{split}
\end{equation}
while $\lambda_1(\cdot)$ stands for the first eigen-value.

\textsc{\textit{Step 3: Bishop-Gromov inequality\footnote{See e.g. \cite{Bishop1}, \cite{Bishop2} to understand it.}.}} 

Re-call the Riemannian manifold and its charactristics discussed in \textit{Cheng eigen-value comparison theorem}. Now, \textit{Bishop-Gromov inequality} indicates that 
\begin{equation}
\begin{split}
 \Gamma(\mathscr{r})=\frac{ \mathbb{V}\mathscr{ol}\Big( \mathscr{B}_{\mathscr{N}(k)}(\mathscr{r})\Big) }{  \mathbb{V}\mathscr{ol}\Big(\mathscr{B}_{\mathscr{M}}(\mathscr{p}, \mathscr{r})\Big)},
\end{split}
\end{equation}
is a non-increasing term/function over $(0, \infty)$.

\textsc{\textit{Step 4.}} Now, the pair $\Big( \eta, \hat{\beta}_{\mathcal{X}} \Big)$ can be a Nash-equilibrium if:
\begin{itemize}
\item (\textit{i}) $ \eta$ is strictly monotone on $\hat{\beta}_{\mathcal{X}} \times \hat{\beta}_{\mathcal{X}}$;
\item (\textit{ii}) the relative sectional curvature is bounded from above.
\end{itemize}

This is because of the fact that the pair $\Big( \eta, \hat{\beta}_{\mathcal{X}} \Big)$ is marginally convex.

The proof is now completed.$\; \; \; \blacksquare$

\section{Proof of Proposition 5}
\label{sec:F}
The proof is provided in terms of the following muti-step solution.

\textsc{\textit{Step 1.}}

If the dimension $n$ is bounded by $\frac{c_1c_2}{c_3}log \frac{2}{c_3}$ and we technically have
\begin{equation}
\begin{split}
 \sum_{i=1}^n\mathscr{Pr}\Big(\mathcal{J}^T \big ( \mathscr{R} \circ y\big )\mathcal{J}>c_4\Big)\ge nc_3,
\end{split}
\end{equation}
while $c_1, c_2, c_3, c_4$ are constant values, then we technically have\footnote{See e.g. \cite{concentraion1}-\cite{concentraion11} for concentration inequalities.} 
\begin{equation}
\begin{split}
 \mathscr{Pr}\bigg( \sigma_m \Big(  \frac{1}{n}\sum_{i=1}^n \big ( \mathscr{R} \circ y\big )_i \Big) \le \frac{c_3c_4}{2} \bigg)\le exp \Big(  -\frac{nc_3}{c_1}   \Big),
\end{split}
\end{equation}
while $\sigma_m$ stands for the singular functions.

\textsc{\textit{Step 2.}}

Foroni has shown that\footnote{See e.g. \cite{concentraion9}, \cite{Foroni1}-\cite{Foroni4} for concentration inequalities.} the eigenvalues of the curvatures are related to the \textit{Lyapunov exponent} which is a quantity through which the rate of separation of infinitesimally close trajectories are actualised, that is, stability and Nash-equilibrium.

The proof is now completed.$\; \; \; \blacksquare$

\section{Proof of Theorem 2}
\label{sec:G}
The proof is provided in terms of the following multi-step solution.

\textsc{\textit{Step 1.}}

We theoretically indicate that\footnote{See e.g. \cite{concentraion1}-\cite{concentraion11} for concentration inequalities.} for any closed set $\mathscr{N} \subset \partial \Omega$ we have 
\begin{equation}
\begin{split}
 \lim_{\epsilon \rightarrow 0} \epsilon \log \mathscr{Pr} \Big(  \underbrace{ \Theta^{\star} \in \mathscr{N} }_{\stackrel{def}{=}  \vert \theta -\bar{\theta} \vert< c_0}   \Big)  \le  \eta(0,t),
\end{split}
\end{equation}
while $c_0$ is a constant.

\textsc{\textit{Step 2.}}

Before we proceed, we should prove the duality between $ \mathscr{P}^{(\mathscr{e})}_{ \pi^{\prime}} \ge \mathscr{Pr} \big \lbrace \pi^{\prime}  \notin \Omega \big \rbrace$ and $\mathscr{Pr} \Big(  \underbrace{ \Theta^{\star} \in \mathscr{N} }_{\stackrel{def}{=}  \vert \theta -\bar{\theta} \vert< c_0}   \Big)$. 

The method used in \textit{Step 3} of Appendix \ref{sec:D} may help us in denoting that 
\begin{equation}
\begin{split}
 \mathscr{Pr}\big \lbrace \pi_i^{\prime}  \in \partial\Omega \big \rbrace \le   c_0 e^{-n \mathop{{\rm \mathbb{I}nf}}\limits_{\mathscr{d}_0 \in \pi_i^{\prime}  \cap \partial \Omega} \mathscr{D}_{Kl} (\mathscr{d}_0 \vert \partial \Omega \setminus \pi_i^{\prime} )},
\end{split}
\end{equation}
holds where $c_0$ is a constant, while we, on the other hand, can write\footnote{See e.g. \cite{concentraion1}-\cite{concentraion11} for concentration inequalities.}
\begin{equation}
\begin{split}
\mathop{{\rm \mathbb{I}nf}}\limits_{ \pi_i^{\prime}  }  \mathscr{Pr}\big \lbrace  \pi_i^{\prime}  \in \partial \Omega \big \rbrace \le c_0  \mathbb{E} \Big( \partial \Omega \setminus \pi_i^{\prime} \Big) ,
\end{split}
\end{equation}
while $c_0$ is a constant, something that gives us, in addition to its previous equation, the following
\begin{equation}
\begin{split}
c_0e^{-n \mathop{{\rm \mathbb{I}nf}}\limits_{\mathscr{d}_0 \in \pi_i^{\prime}  \cap \partial \Omega} \mathscr{D}_{Kl} (\mathscr{d}_0 \vert \partial \Omega \setminus \pi_i^{\prime} )} \cong  \mathbb{E} \Big( \partial \Omega \setminus \pi_i^{\prime} \Big) ,
\end{split}
\end{equation}
while $c_0$ is a constant, something that interpretes for us that the \textit{divergence between the hole we have drilled and the sand we have obtained}, that is, $\hat{\beta}_{\mathcal{X}} (t)$ or correspondingly, the control input $\theta(t)$.

\textsc{\textit{Step 3.}}

Now, $\eta(0, t)$ strongly imposes that $\partial_t \mathcal{U}+\mathop{{\rm \mathbb{I}nf}}\limits_{\theta (\cdot)} {\rm \; } \big \lbrace \hat{\beta}_{\mathcal{X}} (t) -\bar{\theta}(t) + \mathscr{W}_3\big \rbrace+\mathscr{W}_2=0$ is mathematically soften to
\begin{equation}
\begin{split}
 \partial_t \mathcal{U}=\mathop{{\rm \mathbb{I}nf}}\limits_{\theta (\cdot)} \bar{\theta}(t) +\mathscr{W}^{\prime},
\end{split}
\end{equation}
while $\mathscr{W}^{\prime}$ is a random-walk process. This means that we have
\begin{equation}
\begin{split}
 \mathcal{U}=\int_{t}\Big(\mathop{{\rm \mathbb{I}nf}}\limits_{\theta (\cdot)} \bar{\theta}(\mathscr{v}) +\mathscr{W}^{\prime}(\mathscr{v})  \Big) d\mathscr{v},
\end{split}
\end{equation}

\textsc{\textit{Step 4.}}

Now, the \textit{$L_1-$error in kernel density estimation}\footnote{See e.g. \cite{concentraion1}-\cite{concentraion11} for concentration inequalities.} technically indicate that if we have
\begin{equation}
\begin{split}
\mathscr{Z}_n\stackrel{def}{=} \Big \vert \mathcal{U}\big( \theta_1, \cdots, \theta_i, \cdots, \theta_n  \big)- \mathcal{U}\big( \theta_1, \cdots, \theta^{\prime}_i, \cdots, \theta_n  \big)\Big \vert\\ \le \frac{1}{n}\int_{t}\Big(\mathop{{\rm \mathbb{I}nf}}\limits_{\theta (\cdot)} \bar{\theta}(\mathscr{v}) -\mathop{{\rm \mathbb{I}nf}}\limits_{\theta (\cdot)} \bar{\theta}^{\prime}(\mathscr{v}) \Big) d\mathscr{v},
\end{split}
\end{equation}
while $\bar{\theta}^{\prime}(\mathscr{v})$ is arisen due to the essential existence of $\theta^{\prime}_i$ instead of $\theta_i$, then we have
\begin{equation}
\begin{split}
\mathscr{Pr} \bigg(\big \vert \mathscr{Z}_n-\mathbb{E} (\mathscr{Z}_n) \big\vert \ge c_1 \bigg)\le 2 exp \Big(  \frac{-nc_1^{2}}{2}  \Big),
\end{split}
\end{equation}
while $c_1$ is a constant $-$ something which strongly proves that we have
 \begin{equation}
\begin{split}
\mathscr{Pr} \bigg(\big \vert \mathscr{Z}_n-\mathbb{E} (\mathscr{Z}_n) \big\vert \le c_1 \bigg)\ge 1-2 exp \Big(  \frac{-nc_1^{2}}{2}  \Big),
\end{split}
\end{equation}
or correspondingly, with a probability of at least $1-2 exp \Big(  \frac{-nc_1^{2}}{2}  \Big)$ we have $\mathscr{Pr} \bigg(\big \vert \mathscr{Z}_n-\mathbb{E} (\mathscr{Z}_n) \big\vert \le c_1 \bigg)$ or equivalently $\mathscr{Pr} \bigg(\big \vert \theta -\bar{\theta} \vert< c_1 \bigg)    $.

\textsc{\textit{Step 5.}}

Now, with regard to $\partial _t \mathcal{M}\big(t; \hat{\beta}_{\mathcal{X}}(t)\big)= \partial _{\hat{\beta}_{\mathcal{X}}} \bigg( \underbrace{\theta(t;\hat{\beta}_{\mathcal{X}}) \mathcal{M}\big(t; \hat{\beta}_{\mathcal{X}}(t)\big)}_{\stackrel{def}{=} \mathscr{Q}_n} \bigg)+\mathscr{W}_4$, the random-walk Winner process $\mathscr{W}_4$ imposes us to denote the following $-$ as the \textit{concentration-inequality for randomly weighted sum}\footnote{See e.g. \cite{concentraion1}-\cite{concentraion11} for concentration inequalities.}
\begin{equation}
\begin{split}
\mathscr{Pr} \bigg(\big \vert \mathscr{Q}_n-\mathbb{E} (\mathscr{Q}_n) \big\vert \ge c_2 \bigg)\le 2 exp \Big(  \frac{-c_2^{2}}{2 \mathop{{\rm max}}\limits_{n}\vert\vert \mathcal{M}_n \vert\vert_2^2}  \Big),
\end{split}
\end{equation}
while $c_2$ is a constant $-$ something which strongly proves that we have
 \begin{equation}
\begin{split}
\mathscr{Pr} \bigg(\big \vert \mathscr{Q}_n-\mathbb{E} (\mathscr{Q}_n) \big\vert \le c_2 \bigg)\ge 1-2 exp \Big(  \frac{-c_2^{2}}{2\mathop{{\rm max}}\limits_{n} \vert\vert \mathcal{M}_n \vert\vert_2^2}  \Big),
\end{split}
\end{equation}
or correspondingly, with a probability of at least $1-2 exp \Big(  \frac{-c_2^{2}}{2 \mathop{{\rm max}}\limits_{n} \vert\vert \mathcal{M}_n \vert\vert_2^2}  \Big)$ we have $\mathscr{Pr} \bigg(\big \vert \mathscr{Q}_n-\mathbb{E} (\mathscr{Q}_n) \big\vert \le c_2 \bigg)$.

\textsc{\textit{Step 6.}} 

To provisionally conclude, Algorithm \ref{fff} converges for $\mathscr{P}_1$ with probability at least 
\begin{equation}
\begin{split}
\Bigg(1-2 exp \Big(  \frac{-nc_1^{2}}{2}  \Big) \Bigg) \Bigg(  1-2 exp \Big(  \frac{-c_2^{2}}{2 \mathop{{\rm max}}\limits_{n} \vert\vert \mathcal{M}_n \vert\vert_2^2}  \Big) \Bigg).
\end{split}
\end{equation}

\textsc{\textit{Step 7.}} 

In relation to the term $\mathscr{Pr} \Bigg( \theta_{\mathscr{Eve}} \big(v; \hat{\beta}_{\mathcal{X}} \big)=0\bigg \vert \int_0^{v^{-}}\frac{d\theta_{\mathscr{Eve}} \big(v^{\prime}; \hat{\beta}_{\mathcal{X}} \big)}{dv^{\prime}}dv^{\prime}\neq 0\Bigg)dv$ from the relative control-law and according to Appendix \ref{sec:D}, we call 
\begin{equation}
\begin{split}
\underbrace{\theta_{\mathscr{Eve}} \big(v^+; \hat{\beta}_{\mathcal{X}} \big)}_{=0} \le \underbrace{\theta_{\mathscr{Eve}} \big(v; \hat{\beta}_{\mathcal{X}} \big)}_{=0} \le \underbrace{ \theta_{\mathscr{Eve}} \big(v^-; \hat{\beta}_{\mathcal{X}} \big)}_{\neq 0}.
\end{split}
\end{equation}
Now, we can denote\footnote{See e.g. \cite{concentraion1}-\cite{concentraion11} for concentration inequalities.}
\begin{equation}
\begin{split}
\mathbb{E} \Big( e^{\theta_{\mathscr{Eve}} \big(v; \hat{\beta}_{\mathcal{X}} \big)}  \Big) \le \;\;\;\;\;\;\;\;\;\;\;\;\;\;\;\;\;\;\;\;\;\;\;\;\;\;\;\;\;\;\;\;\;\;\;\; \\ exp \bigg(  \frac{1}{8}\Big( \theta_{\mathscr{Eve}} \big(v^-; \hat{\beta}_{\mathcal{X}} \big)-\theta_{\mathscr{Eve}} \big(v^+; \hat{\beta}_{\mathcal{X}} \big)  \Big)^2  \bigg).
\end{split}
\end{equation}

\textsc{\textit{Step 8.}} 

Now, we can write\footnote{See e.g. \cite{eigen1}-\cite{eigen3} for duality confirmation of the maximum eigen-values standing for the SOP.}
\begin{equation}
\begin{split}
\mathscr{Pr} \bigg(  \lambda^{\Big(\theta_{\mathscr{Eve}} \big(v^+; \hat{\beta}_{\mathcal{X}} \big)\Big)}_{max} \le c_3 \bigg) \ge \; \; \; \; \; \; \; \; \; \; \; \; \; \; \;\;\;\;\;\;\;\;\;\;\;\;\;\;\;\;\;\;\\ 1-exp \bigg(  \frac{1}{8}\Big( \theta_{\mathscr{Eve}} \big(v^-; \hat{\beta}_{\mathcal{X}} \big)-\theta_{\mathscr{Eve}} \big(v^+; \hat{\beta}_{\mathcal{X}} \big)  \Big)^2  -c_3\bigg),
\end{split}
\end{equation}
while $c_3>0$ is a constant $-$ something that reveals if how much probability we do need to guarantee the spectral density of the control input $\theta$, that is, the revisited Algorithm \ref{fff} to solve $\mathscr{P}_2$ converges with probability of at least
\begin{equation}
\begin{split}
\Bigg(1-2 exp \Big(  \frac{-nc_1^{2}}{2}  \Big) \Bigg) \Bigg(  1-2 exp \Big(  \frac{-c_2^{2}}{2 \mathop{{\rm max}}\limits_{n} \vert\vert \mathcal{M}_n \vert\vert_2^2}  \Big) \Bigg) \cdots \;\;\; \\ \times \Bigg(1-exp \bigg(  \frac{1}{8}\Big( \theta_{\mathscr{Eve}} \big(v^-; \hat{\beta}_{\mathcal{X}} \big)-\theta_{\mathscr{Eve}} \big(v^+; \hat{\beta}_{\mathcal{X}} \big)  \Big)^2  -c_3\bigg)\Bigg).
\end{split}
\end{equation}

The proof is now completed.$\; \; \; \blacksquare$

%\begin{IEEEbiography}[{\includegraphics[width=1in,height=1.25in,clip,keepaspectratio]{picture}}]{Makan Zamanipour}
%MAKAN ZAMANIPOUR (Researcher-ID: P-6298-2019; ORCID: 0000-0003-1606-9347; Scopus-ID: 56719734800) IEEE Member since 2015, born in Iran on 1983. His main research-field is Wireless communication theory, Information theory, Game theory and Optimisation. He has published a lot of papers in ISI-indexed journals as wll as reviewing for high-prestige ISI-indexed journals in IEEEs, Elsevier etc. His Google-Scholar profile and Publons are available online.
%\end{IEEEbiography}
\markboth{IEEE, VOL. XX, NO. XX, X 2022}%
{Shell \MakeLowercase{\textit{et al.}}: Bare Demo of IEEEtran.cls for Computer Society Journals}
\end{document}